\documentclass[twocolumn]{aastex702}

\usepackage{graphics,graphicx}
\usepackage{natbib}
\usepackage{epsfig}    
\usepackage{subfigure}

\newcommand\avgmstar{\langle M_* \rangle}

\newcommand{\chandra}{\emph{Chandra}}
\newcommand{\feka}{\hbox{Fe\,K$\alpha$}}
\newcommand{\nh}{\hbox{${N}_{\rm H}$}}
\newcommand{\cmsq}{\hbox{cm$^{-2}$}}

\shorttitle{X-Ray Continuum Region in SDSS J1339+1310}
\shortauthors{Morgan et al.}
\graphicspath{{./}{figures/}}

\begin{document}

\title{The X-Ray Continuum Emission Region in the Lensed Quasar SDSS~J133907.23+131038.6 is Much Smaller than the Accretion Disk}

\correspondingauthor{Christopher W. Morgan}
\email{cmorgan@usna.edu}

\author[0000-0003-2460-9999]{Christopher W. Morgan}
\email{cmorgan@usna.edu}
\affiliation{Volgenau Department of Physics, United States Naval Academy, Annapolis, MD 21402 USA}

\author{James B. Margeson }
\email{jamiemargeson9@gmail.com}
\affiliation{Volgenau Department of Physics, United States Naval Academy, Annapolis, MD 21402 USA}

\author{Gilberto Garcia}
\email{ggarcia@ou.edu}
\affiliation{Homer L. Dodge Department of Physics and Astronomy, The University of Oklahoma, Norman, OK 73019 USA}

\author[0000-0001-9203-2808]{Xinyu Dai}
\email{xdai@ou.edu}
\affiliation{Homer L. Dodge Department of Physics and Astronomy, The University of Oklahoma, Norman, OK 73019 USA}

\author[0000-0003-0110-834X]{Luis J. Goicoechea}
\email{luis.goicoechea@unican.es}
\affiliation{Instituto de F\'{i}sica de Cantabria (CSIC-UC) Avda. de Los Castros s/n, E-39005 Santander, Spain}

\author[0000-0003-3062-7835]{Vyacheslav N. Shalyapin}
\email{vyacheslavshalyapin@gmail.com}
\affiliation{Instituto de F\'{i}sica de Cantabria (CSIC-UC) Avda. de Los Castros s/n, E-39005 Santander, Spain}
\affiliation{O.Ya. Usikov Institute for Radiophysics and Electronics, 12 Acad. Proscury St., UA-61085 Kharkiv, Ukraine}

\author[0000-0003-1697-6596]{George Chartas}
\email{chartasg@cofc.edu}
\affiliation{Department of Physics and Astronomy, College of Charleston, Charleston, SC, 29424, USA}

\begin{abstract}

We analyze microlensing variability in 15 seasons of optical monitoring data and 4 epochs of new X-ray observations of the doubly-imaged gravitationally lensed quasar SDSS
J133907.23+131038.6 to place empirical constraints on the size and structure of that system's X-ray and optical continuum emission regions.  Employing a Bayesian Monte Carlo method,
 we analyzed ground-based optical light curves to constrain the half-light radius of the far-UV source $\log(r_{\rm 1/2, FUV}/{\rm cm})=15.78^{+0.26}_{-0.28}$ at $1930{\rm \AA}$, the rest-
 frame center of the {\it r}-band, assuming a $60^\circ$ inclination angle.  This size corresponds to $\sim100\,{\it r}_{\rm g}$ for a $4.0 \times 10^{8} \: {\rm M_{\odot}}$ black hole. We
  measured the half-light radius of the full band ($0.2-8.0 \: {\rm keV}$) X-ray continuum emission region $\log(r_{\rm 1/2, X_{full}}/{\rm cm})=14.32^{+0.23}_{-0.31}$, a size measurement
   that is consistent with the radius of the innermost stable circular orbit (ISCO)
in the Schwarzschild metric.
Two shifted \feka\ lines caused by microlensing are detected in the stacked spectrum of image A at 5.9 and 8.9~keV at $>99\%$ significance.

\end{abstract}

\keywords{\uat{X-ray quasars}{1821} --- \uat{Quasar microlensing}{1318} --- \uat{Strong gravitational lensing}{1643} --- \uat{Supermassive black holes}{1663}}

\section{Introduction}
\label{sec:intro}

Quasars are the most luminous X-ray sources in the universe, and the aggregated emission from the quasar population
is the dominant contributor to the universal X-ray background.  The details of the physical mechanism that generates these X-rays, however,
remain unclear because the emission structures themselves are impossible to image directly;  quasar
X-ray continuum emission regions are orders of magnitude smaller than the angular resolution limit
of any existing (or planned) X-ray observatory.   Given this limitation, we have been forced to constrain the properties of these sources
with a combination of theoretical modeling and more sophisticated observational techniques.  
Over the decades, many radiative accretion models have been proposed, but surprisingly few generate X-rays.
Of the models that do emit X-rays \cite[e.g.][]{Haardt1991,Haardt1993,Maraschi2003,Hirose2004,Nayakshin2004}, they do so
on very different physical scales relative to the gravitational radius $r_g=GM_{\rm BH}/c^2$ of the black hole.

A first category of models \cite[e.g.][]{Hawley2002}, produce
X-ray emission from a hot corona supported by the UV/optical accretion disk.  These models, hereafter ``disk-corona" models, are of a physical scale ($r \approx 200 \; r_{g}$) similar to the optical accretion disk.
Other models, however, \cite[e.g][]{Martocchia2002,Ghisellini2004} emit the X-ray continuum from a very small region
in the immediate vicinity of the black hole ($r \lesssim 3.0 \; r_{g}$), which, 
depending on the angular momentum of the black hole, may be consistent 
with the radius of the innermost stable circular orbit (ISCO). 
Spectral timing observations with {\it NuSTAR} \cite[e.g.][]{Fabian2015} lend
significant support for models with small continuum emission structures, but the conclusions in this and other similar
investigations are strongly model-dependent. In \cite{Frederick2018}, reverberation mapping was used to constrain the size of the FeK$\alpha$
reflection region in the Seyfert Galaxy 1H 1934-063, but FeK$\alpha$ reflection mapping does
not reveal the structure of the continuum source itself. 
In fact, the only existing quantitative measurements of the sizes of X-ray continuum emission regions have used microlensing
\citep{Morgan2008,Dai2010,Morgan2012,Mosquera2013,Blackburne2014,Macleod2015,Blackburne2015}. In all of these cases, the
X-ray continuum was found to emerge from a region immediately
outside of the black hole's innermost stable circular orbit, ISCO. 

The gravitationally lensed quasar SDSS J133907.23+131038.6 (hereafter SDSS1339) is a doubly-imaged system with source redshift 
$z_s = 2.231$ \citep{Inada2009,Shalyapin2014} and image separation $1\farcs7$ that is lensed by an elliptical galaxy at $z_l = 0.607$ \citep{Goicoechea2016}.
The time delay between the images $\Delta t_{BA} = t_B - t_A = +47^{+5}_{-6} \: {\rm days}$, where image A leads image B 
\citep{Goicoechea2016}, is sufficiently
long that correcting the lightcurves for the delay is necessary to properly discriminate between intrinsic (quasar source) and extrinsic 
(viz. microlensing) variability.  At observed-frame optical wavelengths, 
the system has historically exhibited an impressive $\sim0.5 \; {\rm mags}$ of microlensing variability on several occasions over the 15~years it has been monitored \citep{Goicoechea2016,Shalyapin2021}. This provided one motivation
for studying this system at X-ray wavelengths, because we expected the system to be heavily microlensed at X-ray wavelengths
and both \citet{Popovic2006} and \citet{Jovanovic2008} have demonstrated that the shorter timescale of X-ray
microlensing variability can be exploited to differentiate the structure of the quasar high-energy continuum emission region from the 
accretion disk.   SDSS1339 did not disappoint.  

An equally important motivation for studying this system is that it is known to host an unusually small
optical/UV accretion disk. By analysis of microlensing variability in its optical light curves, \citet{Shalyapin2021} reported that the 
far-UV (FUV) emission $(\lambda_{rest} = 1930 \: {\rm \AA})$ emerges from a region that is only $\sim 40 \; r_{g}$ in size, 
where $r_g = G M_{\rm BH}/c^2$ is the gravitational radius of this system's $4^{+6}_{-2} \times 10^8 \: {\rm M_{\sun}}$ \citep[see][]{Shalyapin2021} black hole .  This is
anomalously small  compared to the predictions of the quasar accretion disk size - black hole mass 
relation \citep{Morgan2010,Morgan2018}, so we were interested to determine if this might be a system in which the 
X-ray emission region is similar to the FUV accretion disk in size.

In Section~\ref{sec:obs}, we present our optical light curves and the new X-ray observations from {\it Chandra}. In Section~\ref{sec:models}, we describe
our macroscopic models of the lens galaxy and its microlensing properties, and in Section~\ref{sec:analysis} we describe our Monte Carlo microlensing
analysis and report the results of our analysis.  In Section~\ref{sec:discussion}, we discuss the implications of these results, and we conclude 
with a summary in Section~\ref{sec:conclusions}.
All calculations and results in this paper assume a flat $\Lambda$CDM cosmology with $\Omega_{\Lambda} = 0.7$, $\Omega_M = 0.3$ and $H_{0}=70 \: {\rm km \: s^{-1} \: Mpc^{-1}}$.

\section{Observations}
\label{sec:obs}

\subsection{Optical Monitoring}
\begin{figure}[b]
\plotone{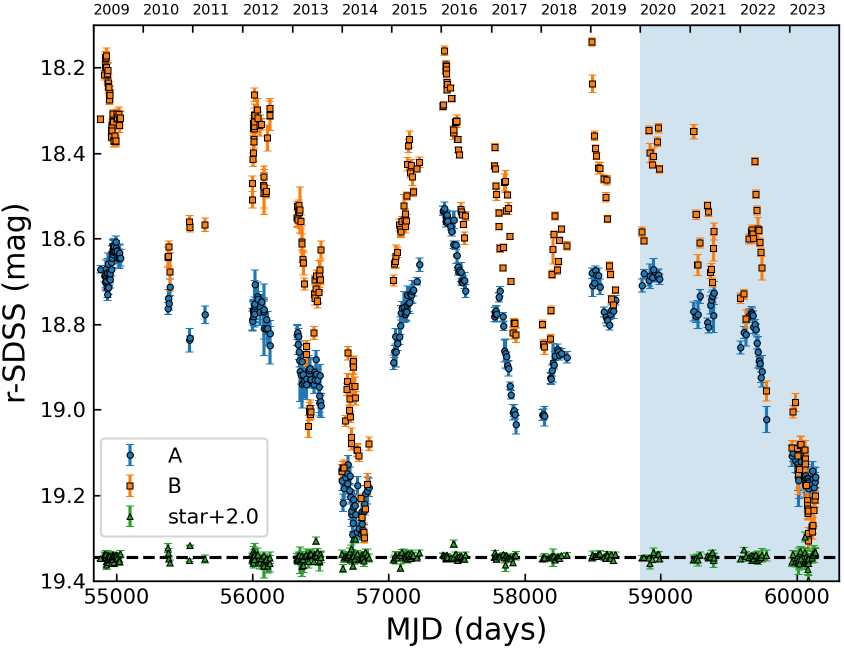}
\caption{Combined {\it r}-band curves of SDSS 1339 based on LT, PS1, MT, ST and NOT observations from 2009 to 2023. Light curves from \citet{Shalyapin2021} correspond to the period 2009$-$2019. The new data
(shaded area) reveal strong quasar variability over the last 4 monitoring years .
\label{fig:opt_lightcurve}}
\end{figure}

SDSS 1339 has been monitored at optical wavelengths by the Gravitational LENses and DArk MAtter (GLENDAMA) project\footnote{https://gravlens.unican.es} since its discovery in 2009 \citep{Inada2009}, and 11 seasons of {\it r}-band data were previously published by \citet{Shalyapin2021}. Although these data were obtained mainly with the 2.0m Liverpool Telescope (LT) in a standard {\it r}-band, the light curves also incorporated some fluxes from Pan-STARRS1 (PS1) {\it r}-band frames, and {\it R}-band frames taken by 1.5m Maidanak Telescope (MT) and the 1.3m SMARTS Telescope (ST). In this paper we add 4 seasons of SDSS {\it r}-band observations from the LT and the 2.56m Nordic Optical Telescope (NOT; exposures taken in April-May 2023) with a typical cadence of 4-10~days.

The 62 epochs of new LT imagery were taken in the SDSS {\it r}-band using the IO:O CCD camera.  Each observational epoch from LT comprises 2 300~s exposures at the $0\farcs30$ pixel scale of IO:O.  Each of the 5 observational epochs from NOT consisted of 1 300~s {\it r}-band image using the NOT/ALOFSC camera at a $0\farcs21$ pixel scale. The details of our photometric pipeline are fully described in \citet{Goicoechea2016} and \citet{Shalyapin2021}, but we will provide a brief summary here.  Since the point-spread function (PSF) of the closely-spaced images ($1\farcs70$)  is blended with flux from the lens galaxy, we employ a photometric fitting technique.  Using IMFITFITS \citep{McLeod1998}, we convolved a deVaucouleurs profile for the lens galaxy with models of the image PSFs based on the measured PSFs of isolated stars in the frame, and we optimized the PSF fit in a given frame by minimizing residuals after model subtraction. Photometric errors were estimated from magnitude changes between adjacent epochs. We calibrated the image magnitudes to the SDSS photometric system using SDSS reference stars in the frames. Extracted NOT light curves were shifted by small offset (-0.039 mag for image A and -0.005 mag for image B) to account for color terms arising from differences in the quantum efficiency curves of the LT and NOT detectors. A plot of the full, compiled light curves is displayed in Figure~\ref{fig:opt_lightcurve}.

\begin{figure}[b]
\plotone{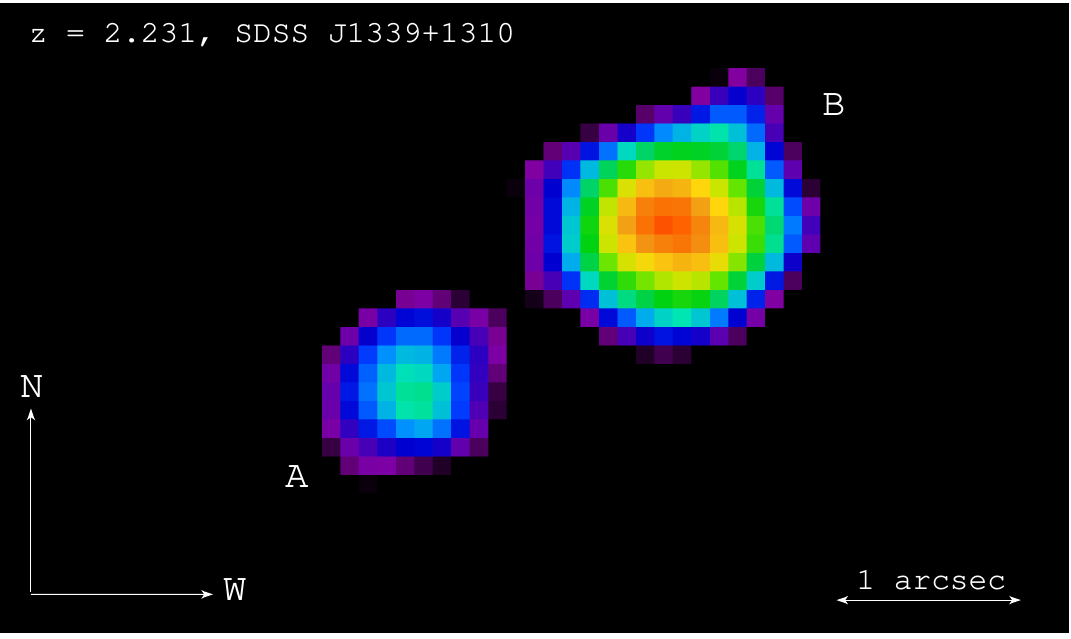}
\caption{Stacked and smoothed full band image of SDSS1339 from \chandra. Total exposure time is 142~ks.
\label{fig:xray}}
\end{figure}
\begin{deluxetable}{lcccc}
        \tabletypesize{\scriptsize}
    \tablecaption{Log of \chandra\ observations of SDSS1339 \label{tab:chandra_obs}}
    \tablewidth{0pt}
    \tablehead{
      \colhead{Sequence} & \colhead{ObsId} & \colhead{Obs.} & \colhead{Exposure} & \colhead{Interval from}  \\
      \colhead{Number} & \colhead{} & \colhead{Date} & \colhead{time} & \colhead{Previous Obs.}  \\
      \colhead{} & \colhead{} & \colhead{(MJD)} & \colhead{(ks)} & \colhead{(days)} 
    }
    \startdata
     704632 & 26711 & 59975.95 & 32.6 & -- \\
     704633 & 26712 & 60028.56 & 15.4 & 53 \\
     704633 & 27756 & 60028.96 & 15.9 & 00 \\
     704634 & 26713 & 60078.73 & 14.9 & 50 \\
     704634 & 27834 & 60092.62 & 17.8 & 14 \\
     704635 & 26714 & 60134.39 & 32.6 & 55 \\
    \enddata
    \tablecomments{Full datset can be found at \dataset[doi:10.25574/cdc.421]{https://doi.org/10.25574/cdc.421}.
    Due to {\it Chandra} scheduling and operational constraints, the second and third epochs were split into two separate observations. ObsIds 26712 and 27756 comprise epoch sequence number 704633, and ObsIds 26713 and 27834 comprise 
    sequence number 704634.  Since ObsIds 26713 and 27834 occurred 14~days apart, we treated the mean of the ObsId 26713 and 27834 dates as the date for ObsId 704634 in our X-ray light curves.}
\end{deluxetable}

\begin{deluxetable*}{lccc}
        \tabletypesize{\scriptsize}
    \caption{X-Ray Spectral Analysis Results \label{tab:spec}}
    \tablewidth{0pt}
    \tablehead{
      \colhead{Model Parameter} & \colhead{Total} & \colhead{A} & \colhead{B}}      
      \startdata
    Power-law index ($\Gamma$)  & $1.83\pm0.04$ & $1.84\pm0.08$ & $1.84\pm0.05$  \\
    Lens \nh\ absorption ($10^{22}$ atoms $\rm cm^{-2}$)  & 0.19 (fixed) & 0.19 (fixed) & 0 (fixed)  \\
    \feka\ line one energy (keV)  & \nodata & $5.90\pm0.12$ & \nodata  \\
    \feka\ line width (keV)  & \nodata & $0.25\pm0.11$ & \nodata  \\
    \feka\ line flux ($10^{-6}$ photons $\rm cm^{-2} \: s^{-1}$)  & \nodata & $3.2\pm1.0$ & \nodata  \\
    \feka\ line equivalent width (keV)  & \nodata & $0.52^{+0.27}_{-0.13}$ & \nodata  \\
    \feka\ line two energy (keV)  & \nodata & $8.94\pm0.11$ & \nodata  \\
    \feka\ line width (keV)  & \nodata & $0.11^{+0.15}_{-0.11}$ & \nodata \\
    \feka\ line flux ($10^{-6}$ photons $\rm cm^{-2} \: s^{-1}$)  & \nodata & $1.7\pm0.7$ & \nodata  \\
    \feka\ line equivalent width (keV)  & \nodata & $0.59\pm0.27$ & \nodata  \\
    \enddata
 \tablecomments{The \feka\ line equivalent width is reported for the rest frame using the XSPEC command equivalent width multiplied by (1+z) to correct for cosmological redshift.}    
\end{deluxetable*}

\subsection{X-Ray Monitoring}
\label{subsec:xray}
We monitored SDSS1339 with four $\sim$35~ksec observations during {\it Chandra} Cycle 24 
with the Advanced CCD Imaging Spectrometer \citep[ACIS][]{Garmire2003}; the target was placed on ACIS-S3 with the TE VFAINT observing mode. 
The second observation contains two separate exposures taken within a single day, and the third observation contains two exposures separated by 14 days.
Specifics of each observation are provided in Table~\ref{tab:chandra_obs}. The observations
were scheduled at $47\pm8$~day intervals corresponding to the measured time delay in this system $\Delta t_{AB} = 47^{+5}_{-6}$ days, where image A leads B 
\citep{Goicoechea2016}, such that the flux ratio at consecutive epochs $f_{A,t_1}/f_{B,t_2}$ represents a time delay corrected contemporaneous measurement of the 
magnification ratio due to strong lensing and microlensing while eliminating contamination from variability intrinsic to the quasar source itself.  
In Figure~\ref{fig:xray} we display a full band, stacked image of SDSS1339 from all \chandra\ observations.

\begin{figure*}
\includegraphics[width=0.50\textwidth]{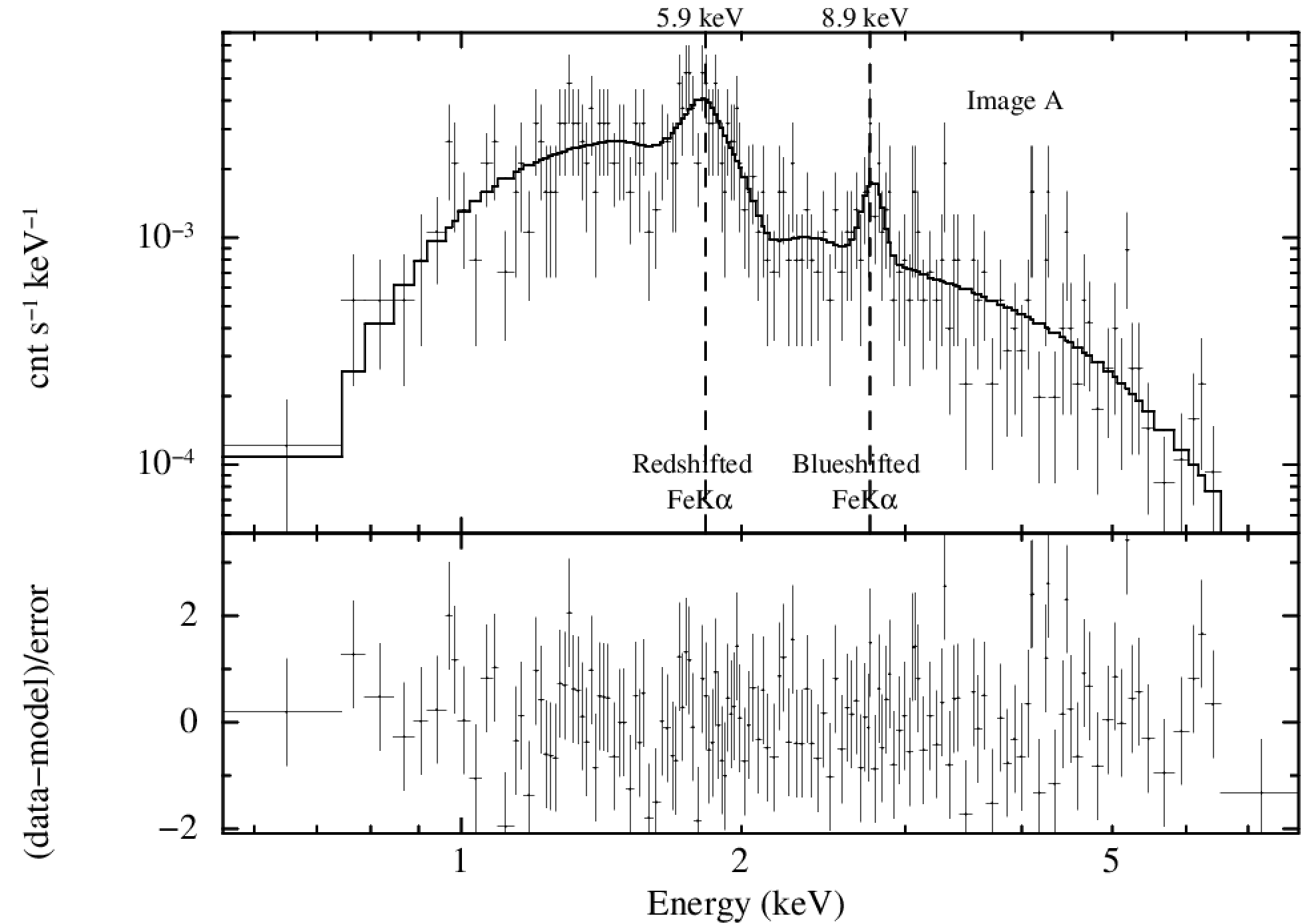}
\includegraphics[width=0.50\textwidth]{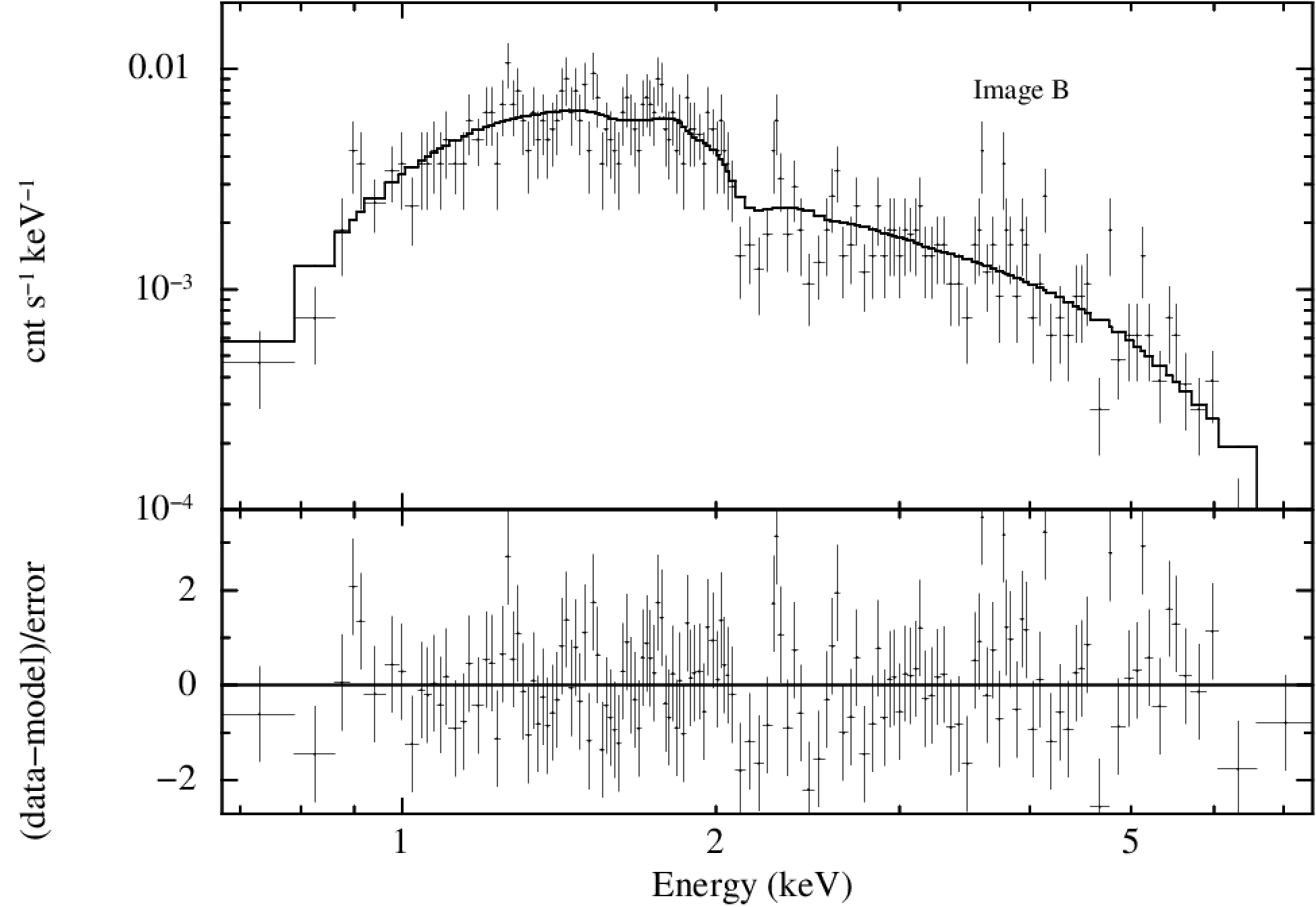}
\caption{X-ray spectra from images A (left) and B (right) fit with a redshifted power-law model modified by Galactic and lens galaxy absorption.  
Two emission lines are detected in the spectrum of image A, modeled by Gaussian profiles.  They are identified as redshifted and blueshifted \feka\ lines at 
$5.90\pm{0.12}$ and $8.94\pm{0.11}$ keV in the quasar restframe.  Residuals from each fit are plotted in the lower panels.  
\label{fig:spec}}
\end{figure*}

We processed all raw data from {\it Chandra} using the CIAO 4.12 software package \citep{CIAO2026}, and then undertook imaging and spectral analyses to measure the flux and spectral properties of the source.
PSF fitting was applied since the two images are blended in the wings of the PSF.  

In the imaging analysis, we fit a PSF model to the lensed components in the full X-ray band (0.2--8~keV) to find the best fit X-ray positions. We constrained the relative positions of the two images using the optical astrometry measurement from \citet{Inada2009}.
After constraining the X-ray image locations, we further measured the flux ratio of the two images in different energy bands.  
We attempted to measure the flux in three energy bands: a soft band, a hard band, and the full energy band that includes both hard and soft photons. However, given the presence of an unusually strong FeK$\alpha$ line at slightly less than 2~keV in the observed frame, we sequestered the photons from the FeK$\alpha$ line into its own energy band, creating three bands in the range 0.2--8.0 keV: a soft band of $0.2-1.6 \: {\rm keV}$, an \feka\ line band of $1.6-2.1 \: {\rm keV}$, and a hard X-ray band of $2.1-8.0 \: {\rm keV}$. We obtain absorbed image count rates from this analysis, which we correct for Galactic and lens galaxy absorption based on the spectral analysis at each epoch.

\begin{figure*}
\includegraphics[width=0.33\textwidth]{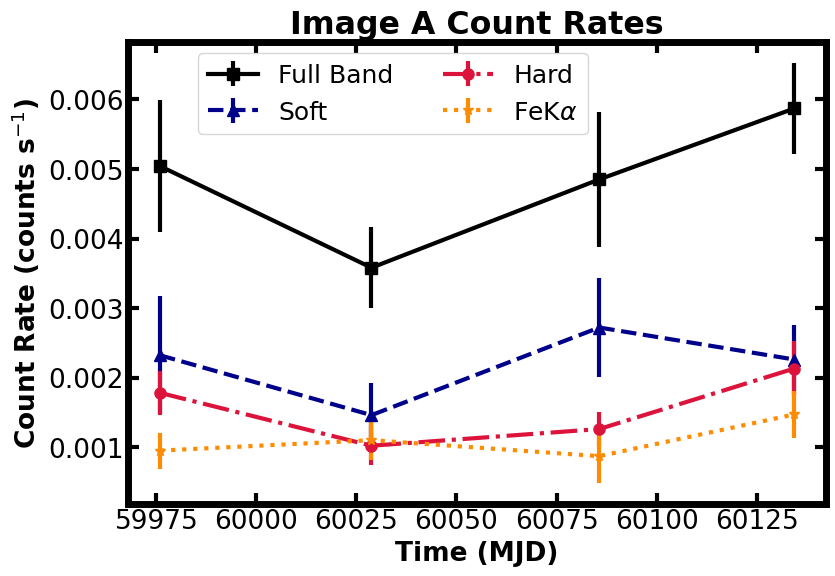}
\includegraphics[width=0.33\textwidth]{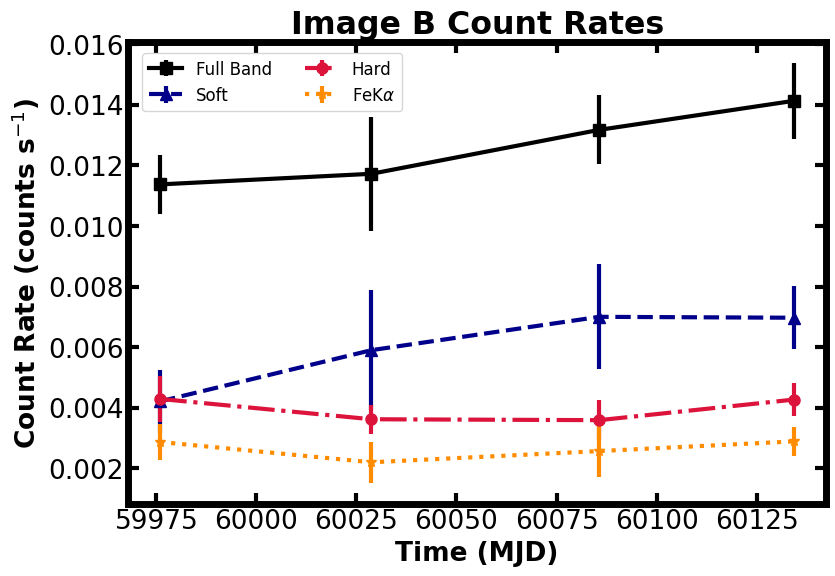}
\includegraphics[width=0.33\textwidth]{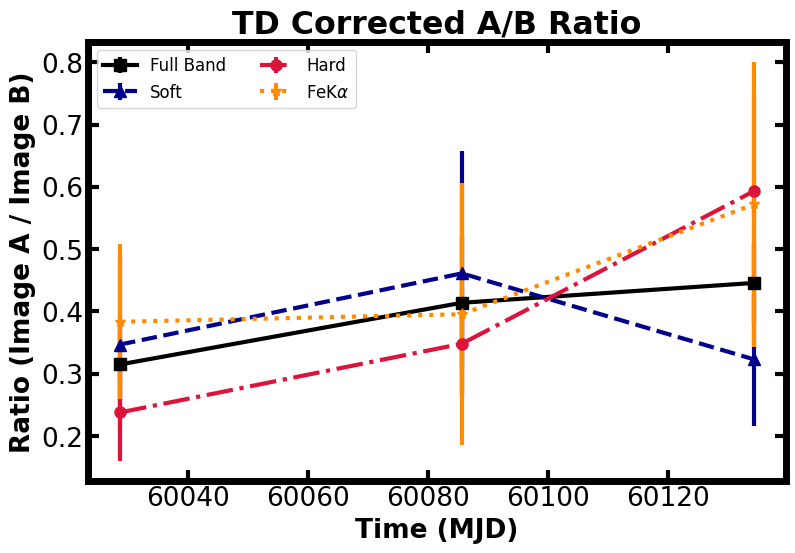}
\caption{X-rays light curves of SDSS1339 image A (left) and B (center) and the time delay corrected flux ratio curve (right) in the full, soft, hard, and \feka\ bands.  Note that the flux in the \feka\ band is a combination of the continuum and \feka\ line emission in the $1.6-2.1 \: {\rm keV}$ energy range. 
\label{fig:xlc}}
\end{figure*}

Individual spectra were extracted for each lensed component using $0\farcs8$ radius circles around the best fit positions. We performed background subtraction using a 
large annulus with an inner radius of 110’’ and an outer radius of 150’’. 
We combined the individual spectra from each epoch to create stacked spectra of each lensed component to better constrain the spectral parameters of the sources.  We performed spectral analysis of the lensed components using XSPEC and c-statistics, with the stacked spectra binned by a range of minimum photon numbers per bin: 1, 3, or 7 photons. 
The different binning yielded consistent spectral fitting results within parameter estimation uncertainties, and we chose to present the fitting results for the 3-count-binning spectrum for image A and 7-count-binning spectrum for image B.
The presented binning yields spectral bin sizes that oversample the $\sim$150~eV spectral resolution of Chandra imaging spectroscopy by $\sim3\times$ in the observed 2--3~keV range.
The stacked spectra were modeled using a redshifted power law (zpowerlw), absorption from the lens galaxy (zphabs), Galactic absorption (phabs), and potential emission lines (zgauss). The Galactic absorption was fixed at $1.58 \times 10^{20} \:\: {\rm atoms}\:\: {\rm cm^{-2}}$ for all components according to \citet{HI4PI2016}. 
Since the effective area of \chandra\ in the soft band has decreased significantly and current \chandra\ spectra are not sensitive to moderate absorption effect, we also fixed the absorption at the lens based on the differential extinction measurement in the system $\Delta E(B-V) = 0.29\pm0.05$ \citep{Goicoechea2016} and the dust-to-gas ratio of cosmic distant galaxies \citep{Dai2009}, yielding a minimum lens absorption of $\nh_A = 1.9\pm0.7 \times 10^{21}~\cmsq$ and $\nh_B = 0~ \cmsq$.  Keeping the lens absorption as free parameters yields consistent fitting values with $\nh_A < 2.3\times10^{21}~\cmsq$ and $\nh_B < 1.2\times10^{21}~\cmsq$.
 The power law photon index and the emission line parameters were allowed to vary during the fitting process.  
   
 \begin{deluxetable*}{lcccccccc}[t]
        \tabletypesize{\scriptsize}
    \tablecaption{Unabsorbed and Deblended \chandra\ Count Rates (ct ${\rm ks}^{-1}$) from SDSS1339 \label{tab:cr}}
    \tablewidth{0pt}
    \tablehead{
      \colhead{Sequence Number} & \multicolumn{2}{c}{\uline{Full Band}} & \multicolumn{2}{c}{\uline{Soft Band}} & \multicolumn{2}{c}{\uline{Hard Band}} & \multicolumn{2}{c}{\uline{Iron Band}}   \\
     \colhead{} & \colhead{A} & \colhead{B} & \colhead{A} & \colhead{B} & \colhead{A} & \colhead{B} & \colhead{A} & \colhead{B}
    }
    \startdata
    704632 & $5.04\pm0.95$ & $11.37\pm0.97$ & $2.32\pm0.85$ & $4.21\pm1.04$ & $1.78\pm0.32$ & $4.29\pm0.75$ & $0.95\pm0.26$ & $2.87\pm0.58$ \\
    704633 & $3.58\pm0.58$ & $11.72\pm1.89$ & $1.46\pm0.46$ & $5.9\pm1.99$ & $1.02\pm0.28$ & $3.62\pm0.48$ & $1.1\pm0.28$ & $2.2\pm0.67$ \\
    704634 & $4.85\pm0.97$ & $13.17\pm1.14$ & $2.72\pm0.71$ & $7.01\pm1.73$ & $1.26\pm0.25$ & $3.59\pm0.67$ & $0.87\pm0.38$ & $2.57\pm0.84$ \\
    704635 & $5.87\pm0.66$ & $14.13\pm1.26$ & $2.26\pm0.49$ & $6.97\pm1.04$ & $2.13\pm0.39$ & $4.27\pm0.53$ & $1.47\pm0.34$ & $2.89\pm0.48$ \\
    \enddata
    \tablecomments{Sequence number 704633 is the combined ObsIds 26712 and 27756. The third epoch is the combined ObsIds 26713 and 27834.} 
\end{deluxetable*}
 
We detected two emission lines in the spectrum of image A at restframe $5.90\pm0.12$ and $8.94\pm0.11$~keV, both at the $>99\%$ significance.
We identify them as shifted \feka\ lines commonly observed in lensed quasars because of the microlensing effects \citep{Chen2012, Chartas2017, Dai2019, Dogruel2020}.
No significant emission lines were detected in image B.
We plot the resulting X-ray spectra of images A and B in Figure~\ref{fig:spec} and provide the details of the best-fit models in Table~\ref{tab:spec}.
We further combined the spectra of images A and B into a total stacked spectrum, and we report the resulting powerlaw index in Table~\ref{tab:spec}.  The same two \feka\ lines remain detectable in the total stacked spectrum at the same energies as reported for image A, so we do not include any
additional line information in Table~\ref{tab:spec}.

Using the power-law index measurements for each epoch derived from the stacked spectra and the assumed Galactic and lens galaxy absorption, we calculated the absorbed to unabsorbed count rate ratios and applied them to the absorbed image count rates from the PSF fitting procedure. We report the unabsorbed and PSF-deblended count rates of the two images in Table~\ref{tab:cr} in the full, soft, hard, and FeK$\alpha$ bands, and we display the  
resulting X-ray light curves in Figure~\ref{fig:xlc}.  A time delay-corrected flux ratio 
curve is also plotted Figure~\ref{fig:xlc}.

Since the sources of hard and soft X-rays are likely to have different sizes, 
we might expect to see some degree of anti-correlation between the hard and soft 
X-ray light curves because smaller sources would be more heavily microlensed.  
To test for this, we calculated the correlation coefficient $r$ between the hard and soft X-ray light curves for each image (A \& B), finding 
$r_{\rm A} = +0.24 \pm 0.46$ and $r_{\rm B} = -0.12 \pm 0.56$.  So, while the soft and hard X-ray light curves are positively correlated in the 
weakly microlensed image A and negatively correlated in the strongly lensed image B, both $r$ values are formally consistent with zero.  
Any existing anti-correlation has been masked by the short, low $S/N$ nature of the single band lightcurves.
 
\section{Lensing Models}
\label{sec:models}

\subsection{Strong Lensing Models}
\label{subsec:stronglens}
Our lens galaxy models are based on the photometric fit to deep {\it I}-band ground based 
imaging as reported by \citet{Shalyapin2014}. These models were first used and are fully described in \citet{Shalyapin2021}, so we provide a brief summary here. 
 Constrained by the quasar image astrometry and the parameters of a deVaucouleurs profile fit to the lens galaxy (position, effective radius, ellipticity and
 ellipticity position angle), we used the LENSMODEL software package \citep{Keeton2001} to generate a series of ten models, consisting of concentric de Vaucouleurs
 and NFW \citep{Navarro1996} profiles, in which the fraction of the total mass in a constant mass-light ratio model $F_{M/L}$ is varied in uniform 
 steps between 0.1 (mostly dark matter) and 1.0 (no dark matter). The resulting total convergence $\kappa$, shear $\gamma$, shear position angle $\theta$ and
  stellar convergence fraction $\kappa_*/\kappa$ are summarized in Table 5 of \citet{Shalyapin2021} and are used as the basis for our lens galaxy microlensing models 
  as described in Section~\ref{subsec:microlensmodels}.

 \subsection{Lens Galaxy Microlensing Models}
 \label{subsec:microlensmodels}
Using the parameters from each of the models in our macro sequence (Section \ref{subsec:stronglens}), we generated magnification patterns 
with the inverse ray shooting technique of \citet{Wambsganss1992} to represent
the set of physically plausible microlensing magnification conditions that could exist along the line of sight through the lens 
galaxy at the location of each lensed image. Constrained by the stellar convergence fraction of the macro model $\kappa_*/\kappa$,
each magnification pattern is populated with a set of stars drawn from the Galactic Initial Mass Function (IMF)\citep{Gould2000}
at random positions. To eliminate possible systematics from the uniqueness of each realization, we generated 40 magnification 
patterns per macroscopic mass model at the location of each image, for a total of 800 magnification patterns. Because our mass 
models cover a wide range of the mass distribution in our lensing galaxy, these patterns span the plausible range of
conditions we would expect to see from the area around images A and B.

When projected onto the source plane, the $8192 \times 8192$ magnification patterns are 40 Einstein radii, $\theta_E$, on a side.
This translates to a physical size on the source plane of
$R_{E} = D_{OS} \; \theta_{E} = 3.3 \times 10^{16} \langle M_* / M_{\odot} \rangle^{1/2} \: {\rm cm} \approx 500 \: r_{g}$
in which $D_{OS}$ is the angular diameter distance between the observer and the source, and $\avgmstar$ is the average mass of a
microlens (star) in the lens galaxy.
It is important that the magnification patterns we create are able to simultaneously accommodate both the largest optical source
and the smallest X-ray source we might expect, 
motivating our use of the very large $8192 \times 8192$ $40 \; \theta_E$ magnification patterns because the pixel scale is the smallest practical size permitted in our simulation $1.6 \times 10^{14} \langle M_* / M_{\odot} \rangle^{1/2}\:{\rm cm}$.

\section{Monte Carlo Analysis and Results}
\label{sec:analysis}

Our Monte Carlo microlensing analysis is based on the method of \citet{Kochanek2004}.  
The technique was fully described in that paper, so
we will only provide a brief summary here. According to Bayes' theorem, given a set of plausible priors on a set of 
physical parameters, the combination of those
model parameters which provide a good fit to observations are also more likely to represent the true physical conditions which
produced the observables. 

A single Monte Carlo trial is conducted as follows: A model quasar source of a trial size is moved across a
single set of magnification patterns for images A and B. 
In practice, we accomplished this by convolving 
a source model kernel at the trial source size with the magnification patterns and then running a point source across the convolved 
patterns. For analytical simplicity we modeled the source as a symmetric, 3-component nested, 2D Gaussian profile simulating the properties
of a standard \citet{Shakura1973} thin disk, since \citet{Mortonson2005} and
\citet{Vernardos2019} have shown
that microlensing statistics are only sensitive to the half-light radius $r_{1/2}$ of the source, not the details of its emissivity profile. The trajectory across the pattern $\phi$ is chosen randomly, and the transverse velocity $v_e$ is 
selected randomly from a uniform logarithmic prior on the transverse 
velocity $10 \: {\rm km \: s^{-1}} \le \hat{v}_e \le 10^5 \: {\rm km \: s^{-1}}$, where $\hat{v}_e$ is in``Einstein
Units" which can be converted to true physical units $v_e$ once the average microlens mass $\avgmstar$ is constrained since
$v_e = \hat{v}_e \: \langle M_*/M_{\odot} \rangle^{1/2}$. We chose a uniform logarithmic prior on $\hat{v}_e$ because a uniform linear prior would bias
the results in favor of high-velocity solutions. 
A small trial source is likely to yield a trial light curve whose magnification changes with greater 
amplitude on shorter timescales, and a larger trial source is more likely to yield a much smoother trial light curve. 

\subsection{Analysis of Optical Light Curves}
\label{subsec:optical}

To constrain the sizes of the FUV and X-ray sources in SDSS1339, we began by 
systematically testing trial FUV quasar source sizes in the range  $14.25 \le \log \left( \hat{r}_s / {\rm cm} \right) \le 17.0$, where the scale radius
in Einstein units $\hat{r}_s$
is related to the scale radius in physical units by $r_s = \hat{r}_s \: \langle M_*/M_{\odot} \rangle^{1/2}$ .  
Similar to our treatment of the effective velocity $\hat{v}_{\rm e}$ parameter, we selected the trial source sizes in uniform logarithmic intervals to avoid biasing the
posteriors in favor of large sizes.
The Gaussian scale radius $r_s$
can easily be converted to a half-light radius for comparisons with other studies $r_{1/2} =2.44 \: r_s$. 

\begin{figure}[t]
\plotone{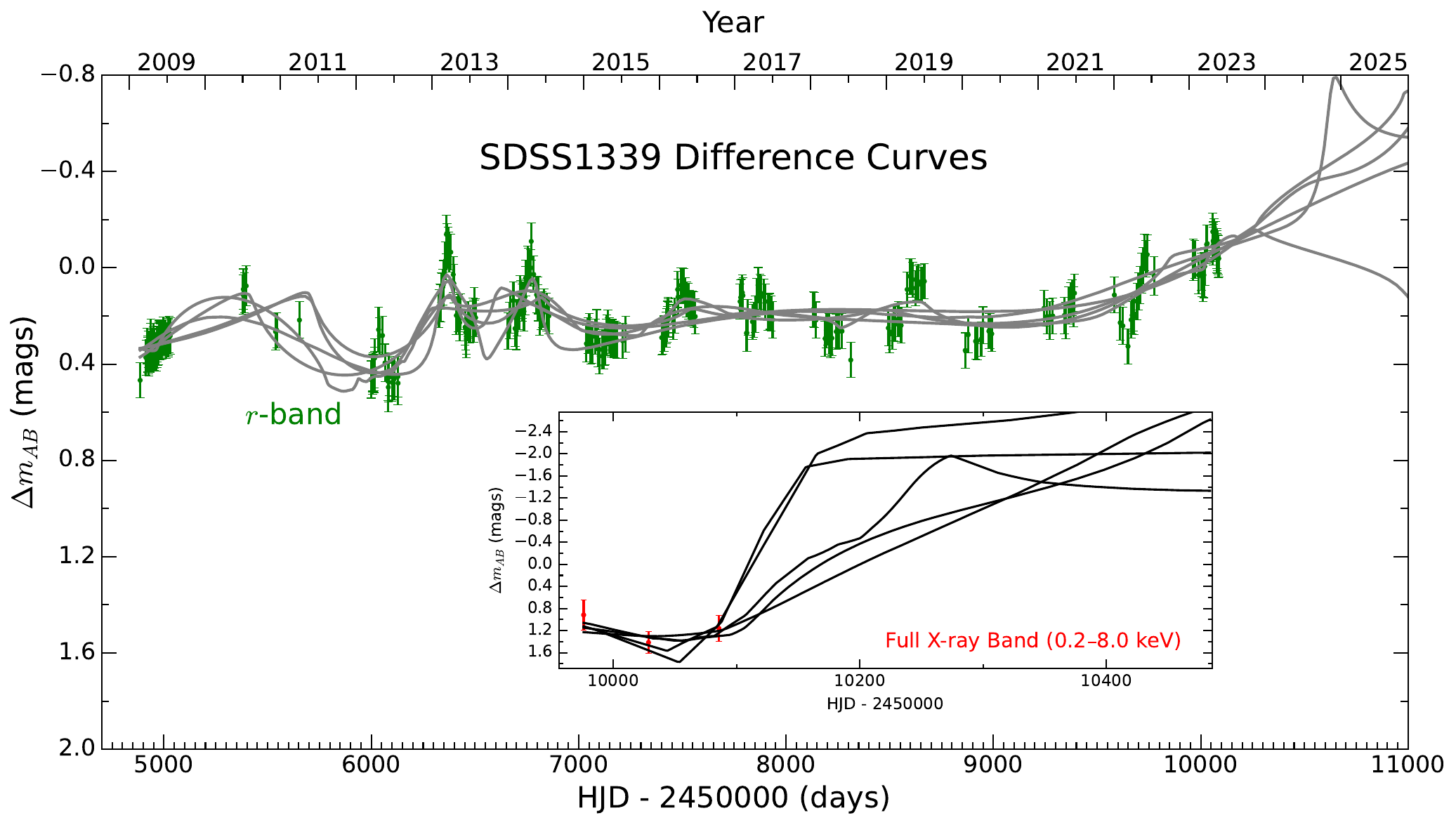}
\caption{{\small Plots of the five best successful joint trials from the joint Monte Carlo analysis of optical and full X-ray band light curves.  
Optical light curve data are plotted in green an the five best trial light curves are shown in gray. X-ray data are plotted in red, and the five best trial X-ray fits are plotted in black. These ``difference curve" plots are found by shifting the light curve for image B by the observed time delays and dividing the flux of image A by flux of B at each epoch. The optical light curves extend from earliest observations in 2009 to 2023, covering 15 seasons.}}
\label{fig:bestcurves_joint}
\end{figure}

As described in Section~\ref{subsec:microlensmodels}, for each macro model we generated 40 independent sets of magnification patterns, and 
for each set of magnification patterns, we attempted $10^8$ fits with our Monte Carlo code for a grand total of $4 \times 10^{10}$ attempted fits to
the optical light curves.  Each set of trial light curves is 
compared to the observed light curves and evaluated with the $\chi^2$ metric. 
The computation was performed on the US Naval Academy (USNA) High Performance Computing Cluster\footnote{https://www.usna.edu/ARCS/hardware/maincluster.php}, where for computational efficiency we aborted all trials with a reduced chi-square $\chi^2/N_{\rm dof} \ge 3.1$ since they 
did not contribute significant statistical weight to the Bayesian integrals in the final analysis. Approximately $2 \times 10^6$ of the $4 \times 10^{10}$ 
trials met our $\chi^2/N_{\rm dof}$ threshold.  The five best fits to the 15 seasons of optical monitoring data are displayed in 
Figure~\ref{fig:bestcurves_joint}, where we have plotted the observations and fits as ``difference light curves" in which the light curve from image A is 
divided by the image B light curve after it was shifted by the system's $47\pm5$~day time delay such that only the extrinsic
variability (viz. microlensing) remains.

\begin{deluxetable*}{lcc}
        \tabletypesize{\scriptsize}
    \tablecaption{Monte Carlo Model Parameters ($\xi$)}
    \tablehead{
      \colhead{Parameter} & {Range} & {Prior} 
    }
    \startdata
    Trial Effective Velocity ($\hat{v}_{\rm e}$) & $10 \le (\hat{v}_{\rm e}/ {\rm km\; s^{-1}}) \le 10^5$ & uniform logarithmic  \\
    Trajectory ($\phi$) & $0^\circ \le \phi < 360^\circ$ (E of N) & uniform linear \\
    Stellar Mass Component Fraction ($F_{M/L}$) & $0.1 \le F_{M/L} \le 1.0 $ & uniform linear \\
    FUV Scale Radius ($\hat{r}_{\rm s,FUV}$) &  $ 14.25 \le \log(\hat{r}_{\rm s,FUV}/{\rm cm}) \le 17.0 $ & uniform logarithmic \\
    X-Ray Scale Radius ($\hat{r}_{\rm s,X}$) & $ 13.2 \le \log(\hat{r}_{\rm s,X}/{\rm cm}) \le 17.0 $ & uniform logarithmic \\
    \hline
    $^*$Effective Velocity Model ($v_{\rm e}$) & see Fig.~\ref{fig:ve} & logarithmic probability density \\
    $^*$Average Microlens Mass $\avgmstar$ & $0.1 \le (\avgmstar/{\rm M_\odot}) \le 1.0$ & uniform logarithmic \\   
    \enddata
    \tablecomments{*Results are convolved with {\it either} the model for the true effective velocity $v_{\rm e}$ or the uniform prior on 
    the average microlens mass $\avgmstar$, but not both.}
    \label{tab:parameters}
\end{deluxetable*}

\subsection{Analysis of X-Ray Monitoring Data}

Because the X-ray light curves are very short (3 epochs after shifting by the time delay), an independent Monte Carlo analysis will not successfully
converge on a meaningful result.  We were, however, very successful in applying a joint analysis requiring a given Monte Carlo trial trajectory to 
fit both the optical and X-ray light curves simultaneously.  In practice, we accomplished this in two phases. In phase 1 (Section~\ref{subsec:optical}),
we found $\sim2 \times 10^6$ successful fits to the optical light curves, and we saved the parameters of those successful trials (e.g. trajectory $\phi$,
effective velocity $\hat{v}_e$ along with their $\chi^2$ score).  In phase 2, we attempted to fit the X-ray data at a range of trial source sizes 
$13.2 \le \log \left( \hat{r}_s / {\rm cm} \right) \le 17.0$ using only the successful trajectories from phase 1, and we evaluated the goodness-of-fit 
for a given trial jointly.

\begin{figure}
\plotone{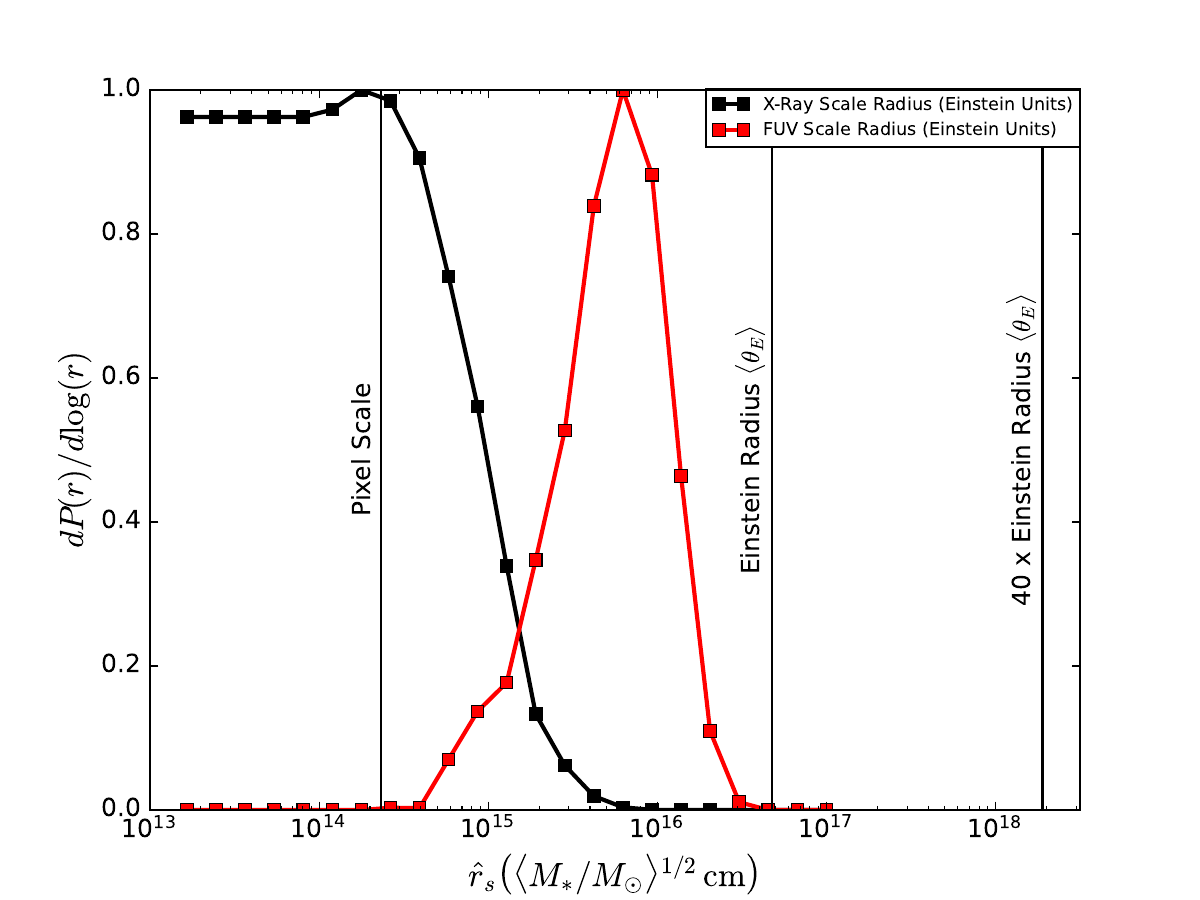}
\caption{{\small Probability density for the scale radius $\hat{r}_s$ of the FUV (red) and X-ray (black) sources in Einstein units based on the joint analysis of the optical and X-ray light curves. The pixel scale, 
the Einstein Radius
$\theta_E$ of a $1{\rm M_\odot}$ star and the $40 \theta_E$ dimension of our magnification patterns are also
plotted.}}
\label{fig:rshat}
\end{figure}

We apply a Bayesian maximum likelihood analysis to the resulting joint set of successful trial 
light curves, yielding probability densities for
parameters of interest.  For example, we find 
the probability density for the source size in Einstein units $\hat{r}_s$ given the results from the Monte Carlo runs $D$  

\begin{equation}
P(\hat{r}_{s}|D) \propto \int P(D|\hat{r}_{s},\xi)\pi(\xi)d\xi
\end{equation} 

by marginalizing over the all of the other parameters of the simulation $\xi$ while applying the
physically-based statistical priors $\pi(\xi)$ on the set of physical parameters.  We display the resulting distribution for
the optical/FUV and X-ray scale radii in Figure~\ref{fig:rshat}.

\begin{figure}
\plotone{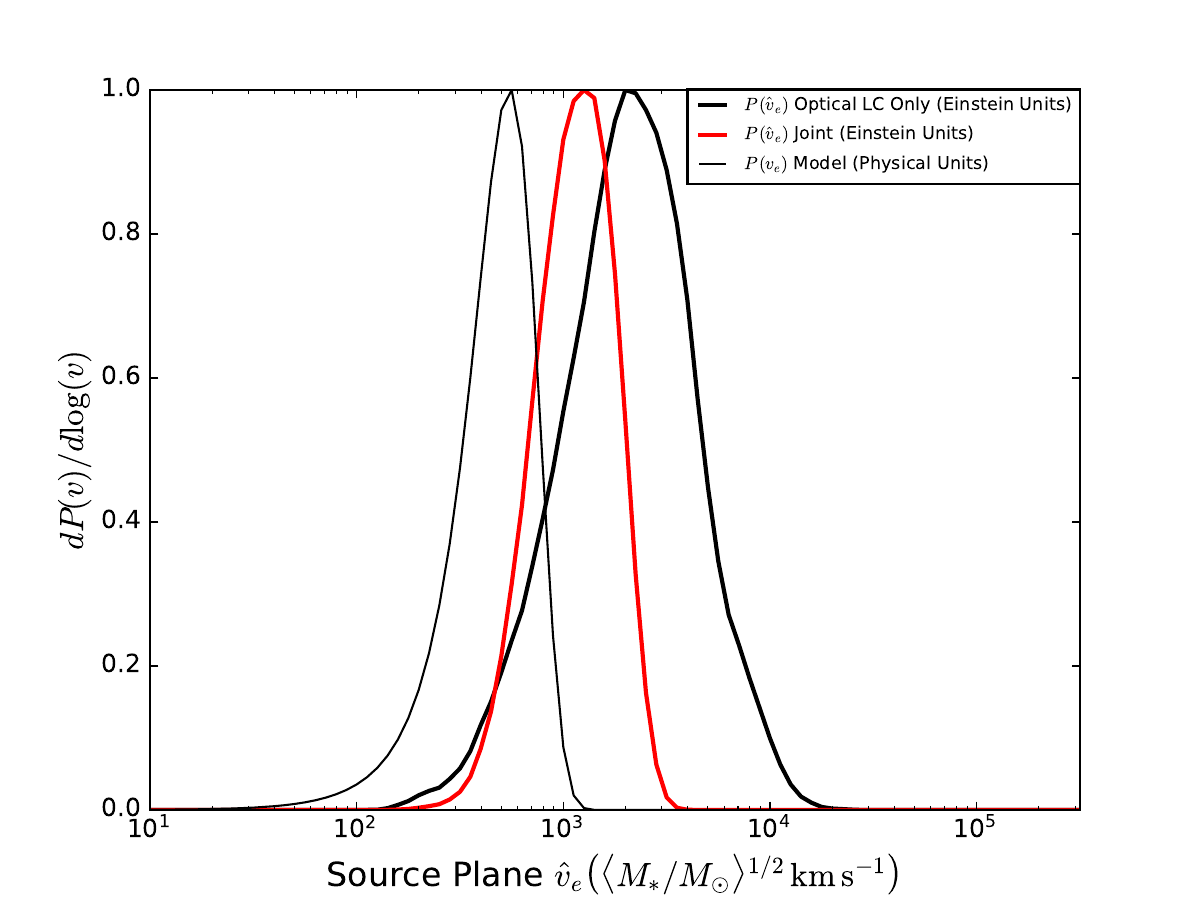}
\caption{{\small Probability density for the transverse velocity of the source in Einstein units $\hat{v}_e$ for the full set of optical light curve solutions (thick black curve) and the joint
X-ray and optical solutions (thick red curve). Our model for
the true physical velocity $v_{\rm e}$ is plotted with a thin black curve.}}
\label{fig:ve}
\end{figure}

Converting the source size from Einstein units into true, physical units
$r_s = \hat{r}_s \: \langle M_*/M_{\odot} \rangle^{1/2}$ requires the application of a physical prior on either the average mass of a lens galaxy star $\avgmstar$
or on the true transverse physical velocity $v_e$.  It is not currently possible to constrain the distribution of $\avgmstar$ in a Local Group galaxy, let 
alone a distant lens galaxy, so any prior we apply on $\avgmstar$ would be speculative.  The probability density for the true effective velocity 
$v_e$, however, can be constrained much more robustly by observations and modeling. 
The velocity across the line of sight $v_e$ is a combination of the contributions from source, lens and observer.
To find the transverse velocity of the observer,
we calculate the cross product of the CMB dipole velocity vector with a unit vector pointing along the line of sight, which yields north and east components
$v_{\rm obs,N} = -110.8 \: {\rm km \: s}^{-1}$ and $v_{\rm obs,E} = -218.5 \: {\rm km \: s}^{-1}$, respectively.  The proper motions of the source and
lens are much too small to measure, so we modeled their peculiar velocities
$\sigma_{pec,s} = 204.1 \: {\rm km \: s}^{-1}$ and $\sigma_{pec,l} = 275.2 \: {\rm km \: s}^{-1}$
 following \citet{Tinker2012} and \citet{Mosquera2011}.  
We estimate the velocity dispersion within the lens galaxy $\sigma_* = 378.5 \: {\rm km \: s}^{-1}$ using a singular isothermal sphere model for the lens. Using Equation 5
of \citet{Mosquera2011}, we calculate a probability density for the transverse velocity in physical units and plot this in Figure~\ref{fig:ve}.

\begin{figure}
\plotone{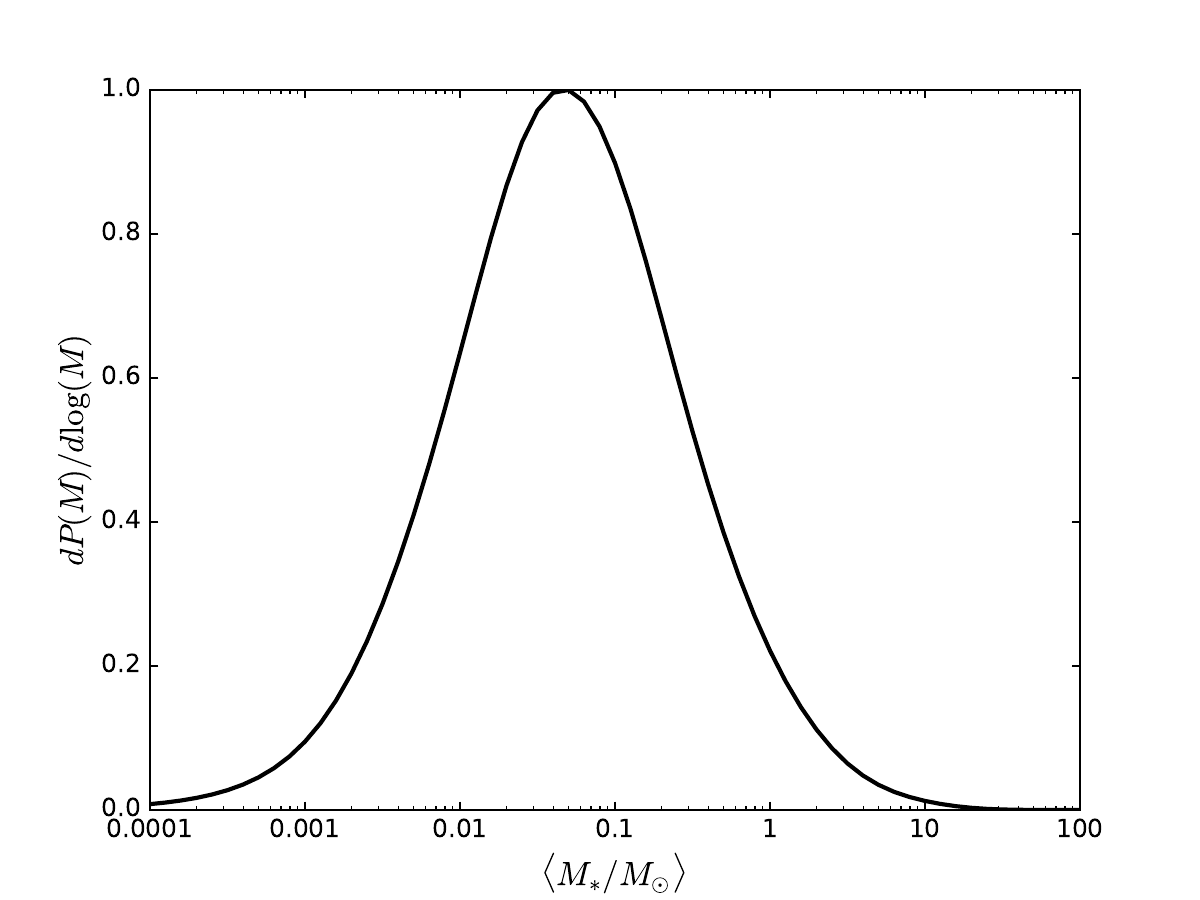}
\caption{{\small Probability density for the average mass of a lens galaxy star $\avgmstar$  for the joint optical/X-ray analysis.}}
\label{fig:mstar}
\end{figure}

\begin{figure}
\plotone{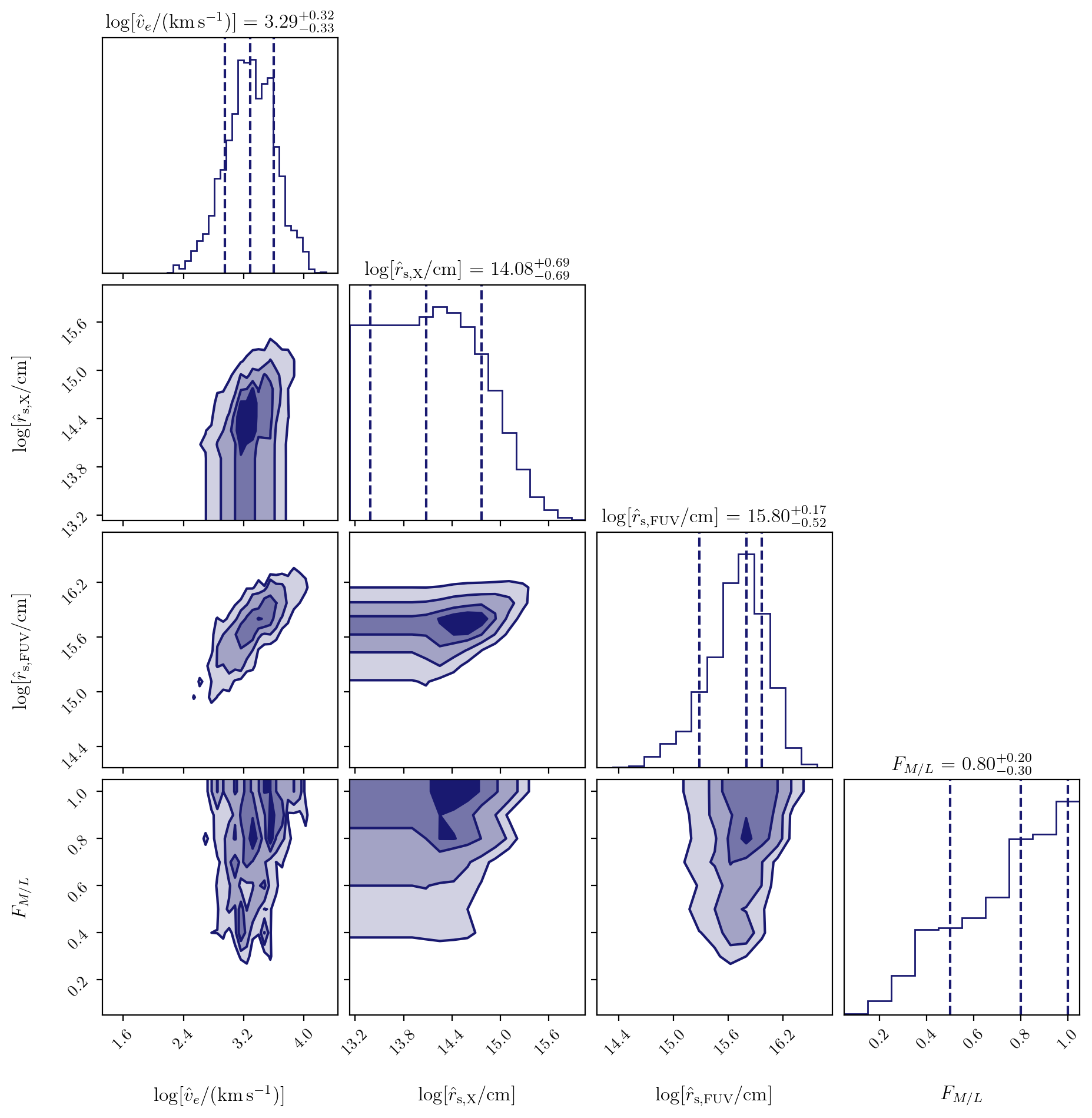}
\caption{{\small Corner plot showing the covariance between the stellar component mass fraction $F_{M/L}$, 
effective velocity in Einstein units $\hat{v}_{\rm e}$, and the scale radius in Einstein units in the rest-frame FUV $\hat{r}_{\rm s, FUV}$ and 
X-rays $\hat{r}_{\rm s,X}$.  Expectation values for these quantities are also provided.}
}
\label{fig:corner}
\end{figure}

\begin{figure}
\plotone{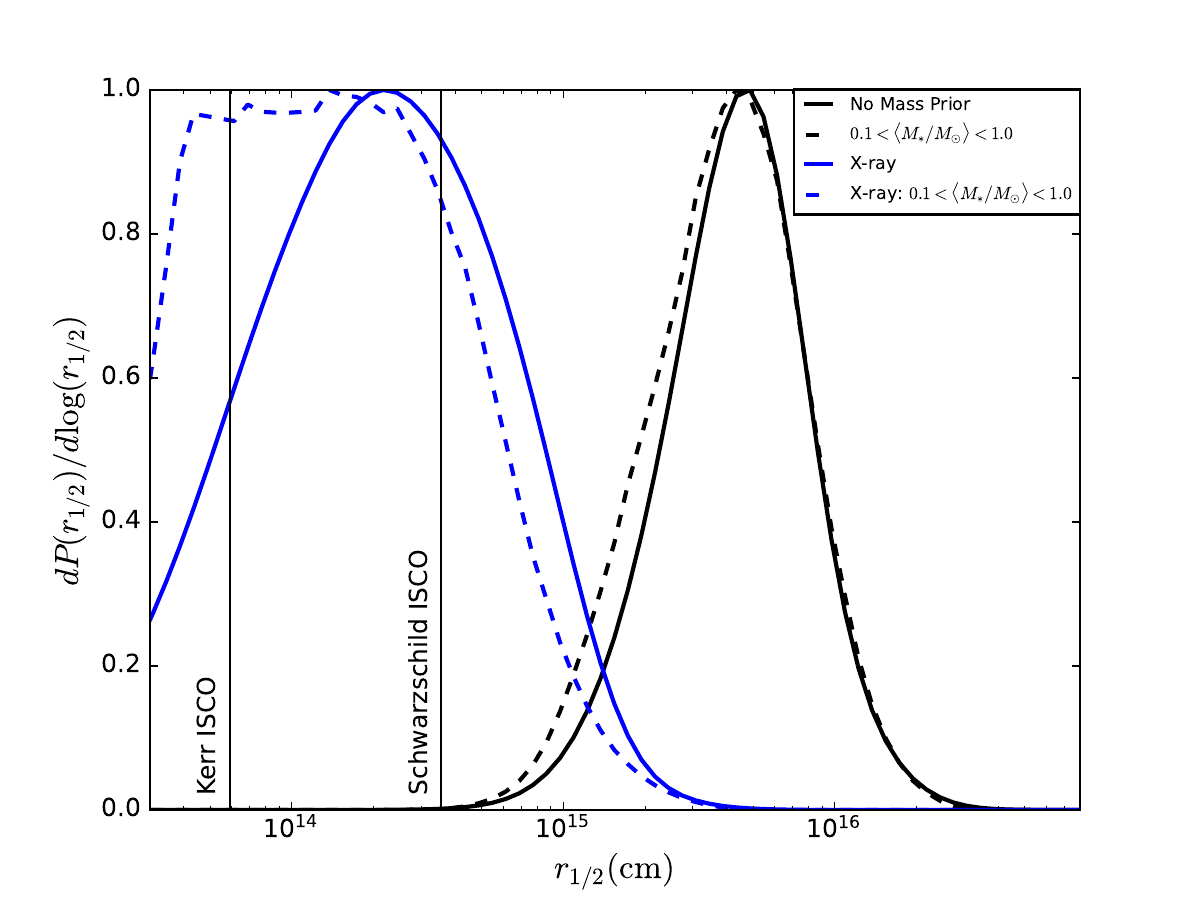} 
\caption{{\small Probability density for the half-light radius of the accretion disk at a rest wavelength $\lambda_{\rm rest} = 1930 \: {\rm \AA}$(black) and the half-light radius of X-ray continuum source (blue)
by analysis of the full 0.2-8.0~keV band. The
solid curves were derived by the convolution of the $\avgmstar$ (Fig.~\ref{fig:mstar}) and $\hat{r}_s$ (Fig.~\ref{fig:rshat}) probability 
distributions, and the dashed curves were derived from $\hat{r}_s$ assuming a uniform prior $0.1 < \langle M_{*}/M_{\odot} \rangle < 1.0$.  
Vertical lines show the innermost stable circular orbit (ISCO) in the Schwarzschild metric at $6 \; r_g$ and the ISCO for a maximally rotating black hole in the Kerr metric at $1 \;r_g$. 
}}
\label{fig:joint_sizes}
\end{figure}

In Figure~\ref{fig:ve}, we also plot the resulting probability density for the effective velocity in Einstein units $\hat{v}_e$, where
since $v_e = \hat{v}_e \: \langle M_*/M_{\odot} \rangle^{1/2}$, we can convolve the $\hat{v}_e$ and $v_e$ curves to produce a probability density for 
the mean stellar mass in the lens galaxy $\avgmstar$, displayed in Figure~\ref{fig:mstar}.  The resulting
constraint on $\avgmstar$ is
very broad, but it did not significantly restrict the precision of our final result. In Figure~\ref{fig:corner}, we provide a corner plot
to demonstrate the covariance between parameters of the simulation.  In simulations of microlensing variability, the effective velocity $\hat{v}_{\rm e}$
and the source size $\hat{r}_{\rm s}$ tend to be somewhat degenerate, nevertheless in this study
FUV scale radius $\hat{r}_{\rm s, FUV}$ converged nicely 
with $\hat{v}_{\rm e}$ due mostly to the strong statistical power of the 14-season optical light curves.  
The X-ray scale radius $\hat{r}_{\rm s,X}$ is weakly bounded on the low end, but, as we describe below, the subsequent 
application of statistical priors on either the true physical velocity $v_{\rm e}$ or the average stellar mass $\avgmstar$ permit a fully bounded 
measurement of the X-ray size when converted to physical units (see Fig.~\ref{fig:joint_sizes}).

The final step in the joint analysis is to convolve the $\avgmstar$ distribution with that of the source sizes in Einstein units $\hat{r}_s$ to yield
the probability density for the source sizes in physical units $r_s$.  We display this result as a half-light radius $r_{1/2}$ in Figure~\ref{fig:joint_sizes},
where as a consistency
check we also plot $r_{1/2}$ using a narrower uniform prior on the average stellar mass $0.1 < \langle M_{*}/M_{\odot} \rangle < 1.0$. 
A summary of
all parameters $\xi$ and statistical priors $\pi(\xi)$ used in our Monte Carlo models is provided in Table~\ref{tab:parameters}.
With the exception of the more sophisticated priors on the true effective velocity $v_e$ and the average microlens mass $\avgmstar$, the parameter
ranges we adopted are very broad, allowing the data to dominate the solution convergence. Tighter parameter ranges can produce narrower posteriors, but
insufficient observational constraints (e.g. high resolution spectroscopy) exist at this time to justify such choices for SDSS1339.

Adopting our results without the uniform prior on the microlens mass, our joint analysis 
yields a half-light radius of $\log(r_{1/2}/{\rm cm})=15.78^{+0.26}_{-0.28}$ at 
a rest-frame wavelength of $\lambda_{\rm rest} = 1930 {\rm \AA}$, assuming 
a $60^\circ$ inclination angle.  It is encouraging that the results using the simplistic prior on the average stellar mass were consistent with the more
robust result derived using the $v_e$ model, but we promote the $v_e$-based measurement as our primary result for this phase of the 
analysis.  This measurement is also consistent with the 11-season result in \citet{Shalyapin2021}.

We began the joint analysis with an attempt to fit the full X-ray band, including all photons from the soft continuum at $0.2-1.6 \: {\rm keV}$, an FeK$\alpha$ band of
 $1.6-2.1 \: {\rm keV}$, and the hard continuum in the range $2.1-8.0 \: {\rm keV}$.  We display the five best joint fits to the optical and 
 X-ray light curves in Figure~\ref{fig:bestcurves_joint}, and the resulting joint constraints on the size of the rest-frame FUV and full-band X-ray 
 continuum emission regions are plotted in Figure~\ref{fig:joint_sizes}.  
 The expectation value for the full X-ray band continuum emission region 
 $\log(r_{\rm 1/2, X_{full}}/{\rm cm})=14.32^{+0.23}_{-0.31}$ is smaller than but formally consistent with the $3.56\times10^{14} \: {\rm cm}$ ISCO in the Schwarzschild metric, and 
 is only $\sim3.5\times$ larger than the $5.9\times10^{13} \: {\rm cm}$ ISCO of a maximally rotating black hole in the Kerr metric. 
 The larger uncertainties on our hard X-ray light curves resulted in a considerably weaker constraint on the hard X-ray continuum source size 
 $\log(r_{\rm 1/2, X_{hard}}/{\rm cm})\leq15.83$, and the soft X-ray constraint was even poorer.
Use of the full band light curves blends photons from the soft and hard X-ray emission sources, 
smoothing the physical scales of what might be two distinct emission regions. Despite this shortcoming, we nevertheless choose to focus only on the stronger 
constraint from the full X-ray band in our interpretation of results in Section~\ref{sec:discussion}, because the low $S/N$ in the soft and hard bands
prevented the Monte Carlo analysis from converging on fully bounded size measurements.
 
 \section{Discussion and Interpretation of Results}
 \label{sec:discussion}
 
The FUV ($\lambda_{\rm rest} = 1930 {\rm \AA}$) continuum emission region is only $\sim100 \: r_{\rm g}$ in size.  While this is consistent with
measurements in a few systems \citep{Dai2010,Blackburne2014,Macleod2015}, FUV accretion disk sizes are much more commonly 
$\gtrsim 200 \: r_{\rm g}$ in size. So, this updated measurement remains among the smallest accretion disk size measurements 
relative to the gravitational radius of the central black hole.   Indeed, it was the anomalously small size of this
system's accretion disk in the optical-only analysis \citep{Shalyapin2021} and its heavily microlensed nature that prompted the proposal for a {\it Chandra} X-ray monitoring program in the first place. 

The ratio between the optical and X-ray region sizes in SDSS1339, $r_{\rm opt}/r_{\rm X,full} \approx 29$ is consistent with earlier microlensing-based
studies \citep[e.g.][]{Macleod2015,Morgan2008,Morgan2012,Mosquera2013} however size ratios in these studies range from $\sim4$ \citep{Mosquera2013} to $\sim50$ \citep{Morgan2008,Morgan2012}. Published studies of quasars RX J1131-1231 \citep{Dai2010} and HE 0435-1223 \citep{Blackburne2014} suggest X-ray region sizes $\la 10 \: r_{\rm g}$ and \citet{Macleod2015} estimated a size of $\sim7\:r_{\rm g}$ for SDSS J0924+0219, but the expectation value for the size of SDSS J1339 is the smallest
to date at approximately $\sim3.5\:r_{\rm g}$.

The ISCO of a black hole is a straightforward (but not algebraically simple) function of the spin parameter $a/M_{BH}$ of the black hole 
(see \citep{Bardeen1972} for a full derivation), where 
in standard GR geometrized form with $c=G=1$, the angular momentum $a$ and $M_{BH}$ both have units of length (cm). In essence, 
as $a/M_{BH} \rightarrow 0$, the ISCO approaches that from the non-rotating Schwarzschild metric $r_{\rm ISCO-Sch} = 6 \: r_g$, and as 
$a/M_{BH}$ approaches its theoretical maximum, $a/M_{BH} \rightarrow 1$, the ISCO approaches $r_{\rm ISCO-Kerr_{max}} = r_g$.  
The constraints we derived on the size of the X-ray continuum emission region are unfortunately too broad to permit a precise measurement of the
angular momentum of the black hole, but given that the expectation value for the X-ray size is $\sim3.5 \; r_g$ we can certainly
say it is \emph{possible} that the SMBH in SDSS1339 has a measureable angular momentum. 

Motivated by this possibility, we are very optimistic about improving the spin constraint in future work.   As discussed in Section~\ref{subsec:xray}, we made robust detections of 
two shifted \feka\ lines, one redshifted to 5.9~keV and one blueshifted to 8.9~keV, in the stacked X-ray spectrum from the fainter image A, but the
$S/N$ was too low to permit a measurement of the ${\rm FeK \alpha}$ line in the spectra from the individual epochs. 
Similar redshifted \feka\ lines at 5.9~keV have been observed in other lens systems such as Q2237+0305 \citep{Dai2003, Chen2012}, SDSS0924+0219 \citep{Chen2012}, RXJ1131$-$1231
\citep{Chartas2017}; however, the blueshifted \feka\ line at 8.9~keV is extremely rare even for lensed quasars. 

It is also interesting to note that in SDSS1339, image B exhibits far more FUV and X-ray continuum microlensing than does image A, yet 
the prominently microlensed \feka\ lines appear in image A, not image B. One possible explanation is that in image B, the compact source of the X-ray continuum 
may be much more magnified than the region of the disk where the
lines are produced by continuum reprocessing. In this interpretation, the \feka\ lines are not resolved because of dilution by the highly-magnified X-ray continuum. Alternatively, some have suggested
that the quasar relativistic X-ray reflection region might be smaller than the X-ray continuum emission region because the equivalent widths of \feka\ emission lines are often much broader in lensed
quasars than they are in unlensed systems \citep[e.g.][]{Chen2012,Dai2019}.  A small X-ray reflection region that is physically separated from the continuum could allow localized caustics from low
mass microlenses in image A to produce both the \feka\ energy shift and magnification without significantly magnifying the larger X-ray continuum or FUV sources \citep[e.g][]{Bhatiani2019}.
Differentiating between these two explanations will require \feka\ line profile modeling in a high $S/N$, multiepoch series of observations.

Analysis of ${\rm FeK \alpha}$ 
emission in a single or stacked epoch can be used to constrain the spin of a black hole
using a relativistic disk reflection model \citep[e.g.][]{Reis2014,Reynolds2014} or from the microlensing magnification effect \citep{Dai2019}.
However, single-epoch studies do not properly account for the influence
of microlensing on the line shape. 
Multiepoch monitoring of the ${\rm FeK \alpha}$ line can yield more robust constraints on the
black hole spin parameter \citep{Chartas2017}. 
Future \chandra\ monitoring of SDSS1339 with greater $S/N$ is needed to realize this objective.

\section{Conclusions}
\label{sec:conclusions}

In this paper, we presented four new epochs of X-ray observations of the doubly-lensed quasar SDSS1339 with {\it Chandra}, which, when analyzed jointly
with 11-season optical light curves from the literature plus 4 new seasons of optical monitoring, yielded joint measurements of the size of the 
rest-frame X-ray and FUV continuum emission regions in this system. The primary results are:
\begin{enumerate}
    \item We measured the scale radius of the FUV source at $\lambda_{\rm rest} = 1930 {\rm \AA}$ $\log(r_{s}/{\rm cm})=15.78^{+0.26}_{-0.28}$ in SDSS1339, corresponding to $~\sim100 r_g$ for the system's black hole.  This is similar to the smallest UV accretion disk sizes measured in other lensed quasars using the multi-epoch light curve analysis technique. 
    \item We measured the half-light radius of the X-ray continuum source $\log(r_{\rm 1/2, X_{full}}/{\rm cm})=14.32^{+0.23}_{-0.31}$ using the full 0.2--8.0 keV X-ray band. Since the expectation value for the full band half-light radius is smaller than the 
    ISCO in the Schwarzschild metric, it is possible that this SMBH may exhibit a detectable signature of rotation.
\end{enumerate}

We continue to monitor SDSS1339 from the ground at optical wavelengths, and we are currently monitoring it in the rest-frame extreme UV (EUV) $(\lambda_{\rm rest} = 827{\rm \AA})$ with {\it HST}. When coupled with the X-ray size measurements from this present work (and the
pending improvements from a proposed future campaign with {\it Chandra}), 
analysis of the 12 epochs of UVIS observations from {\it HST} Cycles 32, 33 \& 34  will complete the most comprehensive picture of the structure of
the high energy continuum source(s) in a single quasar system to date.  Combining the existing accretion disk size measurement at FUV wavelengths with this new EUV size measurement
may also enable an empirical measurement of the temperature profile $T(r)\propto r^{-\beta}$ in this system. Empirical measurements of quasar
accretion disk temperature profiles are of great value to astronomy because wide discrepancies currently exist
between between recent measurements in other systems \citep[e.g.][]{Jimenez2014a,Bate2018,Cornachione2020a,Cornachione2020b} 
and basic thin disk models \citep[e.g.][]{Shakura1973}.  Resolving this discrepancy will be a critical step toward a robust model of quasar radiative accretion
that is consistent with observations.

\begin{acknowledgments}
The scientific results reported in this article are based on
observations made by the Chandra X-ray Observatory.
This paper employs a list of Chandra datasets, obtained by the Chandra X-ray Observatory, contained in the Chandra Data Collection (CDC) 421~\dataset[doi:10.25574/cdc.421]{https://doi.org/10.25574/cdc.421}.
Support for this work was provided by the National Aeronautics and Space Administration through award number GO3-24067Z issued by the
Chandra X-ray Observatory Center, which is operated by the
Smithsonian Astrophysical Observatory for and on behalf of
the National Aeronautics Space Administration under contract
NAS8-03060.
L.J.G. and V.N.S. acknowledge support by Universidad de Cantabria and the grant PID2020-118990GB-I00 funded by MCIN/AEI/10.13039/501100011033.
The views expressed in this article are those of the authors and do not reflect the official policy or position of the U.S. Naval Academy, the Department of the Navy, the Department of Defense, or the U.S. Government.
\end{acknowledgments}

\vspace{5mm}
\facilities{Liverpool Telescope, Nordic Optical Telescope, CXO}
\software{astropy \citep{astropy2013,astropy2018,astropy2022},
	CIAO 4.12 \citep{Fruscione2026},
	lensmodel \citep{Keeton2001}, 
          matplotlib \citep{matplotlib} 
          }


\begin{thebibliography}{}
\expandafter\ifx\csname natexlab\endcsname\relax\def\natexlab#1{#1}\fi
\providecommand{\url}[1]{\href{#1}{#1}}
\providecommand{\dodoi}[1]{doi:~\href{http://doi.org/#1}{\nolinkurl{#1}}}
\providecommand{\doeprint}[1]{\href{http://ascl.net/#1}{\nolinkurl{http://ascl.net/#1}}}
\providecommand{\doarXiv}[1]{\href{https://arxiv.org/abs/#1}{\nolinkurl{https://arxiv.org/abs/#1}}}

\bibitem[{{Astropy Collaboration} {et~al.}(2013){Astropy Collaboration},
  {Robitaille}, {Tollerud}, {Greenfield}, {Droettboom}, {Bray}, {Aldcroft},
  {Davis}, {Ginsburg}, {Price-Whelan}, {Kerzendorf}, {Conley}, {Crighton},
  {Barbary}, {Muna}, {Ferguson}, {Grollier}, {Parikh}, {Nair}, {Unther},
  {Deil}, {Woillez}, {Conseil}, {Kramer}, {Turner}, {Singer}, {Fox}, {Weaver},
  {Zabalza}, {Edwards}, {Azalee Bostroem}, {Burke}, {Casey}, {Crawford},
  {Dencheva}, {Ely}, {Jenness}, {Labrie}, {Lim}, {Pierfederici}, {Pontzen},
  {Ptak}, {Refsdal}, {Servillat}, \& {Streicher}}]{astropy2013}
{Astropy Collaboration}, {Robitaille}, T.~P., {Tollerud}, E.~J., {et~al.} 2013,
  \aap, 558, A33, \dodoi{10.1051/0004-6361/201322068}

\bibitem[{{Astropy Collaboration} {et~al.}(2018){Astropy Collaboration},
  {Price-Whelan}, {Sip{\H{o}}cz}, {G{\"u}nther}, {Lim}, {Crawford}, {Conseil},
  {Shupe}, {Craig}, {Dencheva}, {Ginsburg}, {Vand erPlas}, {Bradley},
  {P{\'e}rez-Su{\'a}rez}, {de Val-Borro}, {Aldcroft}, {Cruz}, {Robitaille},
  {Tollerud}, {Ardelean}, {Babej}, {Bach}, {Bachetti}, {Bakanov}, {Bamford},
  {Barentsen}, {Barmby}, {Baumbach}, {Berry}, {Biscani}, {Boquien}, {Bostroem},
  {Bouma}, {Brammer}, {Bray}, {Breytenbach}, {Buddelmeijer}, {Burke},
  {Calderone}, {Cano Rodr{\'\i}guez}, {Cara}, {Cardoso}, {Cheedella}, {Copin},
  {Corrales}, {Crichton}, {D'Avella}, {Deil}, {Depagne}, {Dietrich}, {Donath},
  {Droettboom}, {Earl}, {Erben}, {Fabbro}, {Ferreira}, {Finethy}, {Fox},
  {Garrison}, {Gibbons}, {Goldstein}, {Gommers}, {Greco}, {Greenfield},
  {Groener}, {Grollier}, {Hagen}, {Hirst}, {Homeier}, {Horton}, {Hosseinzadeh},
  {Hu}, {Hunkeler}, {Ivezi{\'c}}, {Jain}, {Jenness}, {Kanarek}, {Kendrew},
  {Kern}, {Kerzendorf}, {Khvalko}, {King}, {Kirkby}, {Kulkarni}, {Kumar},
  {Lee}, {Lenz}, {Littlefair}, {Ma}, {Macleod}, {Mastropietro}, {McCully},
  {Montagnac}, {Morris}, {Mueller}, {Mumford}, {Muna}, {Murphy}, {Nelson},
  {Nguyen}, {Ninan}, {N{\"o}the}, {Ogaz}, {Oh}, {Parejko}, {Parley}, {Pascual},
  {Patil}, {Patil}, {Plunkett}, {Prochaska}, {Rastogi}, {Reddy Janga},
  {Sabater}, {Sakurikar}, {Seifert}, {Sherbert}, {Sherwood-Taylor}, {Shih},
  {Sick}, {Silbiger}, {Singanamalla}, {Singer}, {Sladen}, {Sooley},
  {Sornarajah}, {Streicher}, {Teuben}, {Thomas}, {Tremblay}, {Turner},
  {Terr{\'o}n}, {van Kerkwijk}, {de la Vega}, {Watkins}, {Weaver}, {Whitmore},
  {Woillez}, {Zabalza}, \& {Astropy Contributors}}]{astropy2018}
{Astropy Collaboration}, {Price-Whelan}, A.~M., {Sip{\H{o}}cz}, B.~M., {et~al.}
  2018, \aj, 156, 123, \dodoi{10.3847/1538-3881/aabc4f}

\bibitem[{{Astropy Collaboration} {et~al.}(2022){Astropy Collaboration},
  {Price-Whelan}, {Lim}, {Earl}, {Starkman}, {Bradley}, {Shupe}, {Patil},
  {Corrales}, {Brasseur}, {N{"o}the}, {Donath}, {Tollerud}, {Morris},
  {Ginsburg}, {Vaher}, {Weaver}, {Tocknell}, {Jamieson}, {van Kerkwijk},
  {Robitaille}, {Merry}, {Bachetti}, {G{"u}nther}, {Aldcroft},
  {Alvarado-Montes}, {Archibald}, {B{'o}di}, {Bapat}, {Barentsen}, {Baz{'a}n},
  {Biswas}, {Boquien}, {Burke}, {Cara}, {Cara}, {Conroy}, {Conseil}, {Craig},
  {Cross}, {Cruz}, {D'Eugenio}, {Dencheva}, {Devillepoix}, {Dietrich},
  {Eigenbrot}, {Erben}, {Ferreira}, {Foreman-Mackey}, {Fox}, {Freij}, {Garg},
  {Geda}, {Glattly}, {Gondhalekar}, {Gordon}, {Grant}, {Greenfield}, {Groener},
  {Guest}, {Gurovich}, {Handberg}, {Hart}, {Hatfield-Dodds}, {Homeier},
  {Hosseinzadeh}, {Jenness}, {Jones}, {Joseph}, {Kalmbach}, {Karamehmetoglu},
  {Ka{l}uszy{'n}ski}, {Kelley}, {Kern}, {Kerzendorf}, {Koch}, {Kulumani},
  {Lee}, {Ly}, {Ma}, {MacBride}, {Maljaars}, {Muna}, {Murphy}, {Norman},
  {O'Steen}, {Oman}, {Pacifici}, {Pascual}, {Pascual-Granado}, {Patil},
  {Perren}, {Pickering}, {Rastogi}, {Roulston}, {Ryan}, {Rykoff}, {Sabater},
  {Sakurikar}, {Salgado}, {Sanghi}, {Saunders}, {Savchenko}, {Schwardt},
  {Seifert-Eckert}, {Shih}, {Jain}, {Shukla}, {Sick}, {Simpson},
  {Singanamalla}, {Singer}, {Singhal}, {Sinha}, {Sip{H{o}}cz}, {Spitler},
  {Stansby}, {Streicher}, {{{S}}umak}, {Swinbank}, {Taranu}, {Tewary},
  {Tremblay}, {Val-Borro}, {Van Kooten}, {Vasovi{'c}}, {Verma}, {de Miranda
  Cardoso}, {Williams}, {Wilson}, {Winkel}, {Wood-Vasey}, {Xue}, {Yoachim},
  {Zhang}, {Zonca}, \& {Astropy Project Contributors}}]{astropy2022}
{Astropy Collaboration}, {Price-Whelan}, A.~M., {Lim}, P.~L., {et~al.} 2022,
  \apj, 935, 167, \dodoi{10.3847/1538-4357/ac7c74}

\bibitem[{{Bardeen} {et~al.}(1972){Bardeen}, {Press}, \&
  {Teukolsky}}]{Bardeen1972}
{Bardeen}, J.~M., {Press}, W.~H., \& {Teukolsky}, S.~A. 1972, \apj, 178, 347,
  \dodoi{10.1086/151796}

\bibitem[{{Bate} {et~al.}(2018){Bate}, {Vernardos}, {O'Dowd}, {Neri-Larios},
  {Webster}, {Floyd}, {Barone-Nugent}, {Labrie}, {King}, \& {Yong}}]{Bate2018}
{Bate}, N.~F., {Vernardos}, G., {O'Dowd}, M.~J., {et~al.} 2018, \mnras, 479,
  4796, \dodoi{10.1093/mnras/sty1793}

\bibitem[{{Bhatiani} {et~al.}(2019){Bhatiani}, {Dai}, \&
  {Guerras}}]{Bhatiani2019}
{Bhatiani}, S., {Dai}, X., \& {Guerras}, E. 2019, \apj, 885, 77,
  \dodoi{10.3847/1538-4357/ab46ac}

\bibitem[{{Blackburne} {et~al.}(2014){Blackburne}, {Kochanek}, {Chen}, {Dai},
  \& {Chartas}}]{Blackburne2014}
{Blackburne}, J.~A., {Kochanek}, C.~S., {Chen}, B., {Dai}, X., \& {Chartas}, G.
  2014, \apj, 789, 125, \dodoi{10.1088/0004-637X/789/2/125}

\bibitem[{{Blackburne} {et~al.}(2015){Blackburne}, {Kochanek}, {Chen}, {Dai},
  \& {Chartas}}]{Blackburne2015}
---. 2015, \apj, 798, 95, \dodoi{10.1088/0004-637X/798/2/95}

\bibitem[{{Chartas} {et~al.}(2017){Chartas}, {Krawczynski}, {Zalesky},
  {Kochanek}, {Dai}, {Morgan}, \& {Mosquera}}]{Chartas2017}
{Chartas}, G., {Krawczynski}, H., {Zalesky}, L., {et~al.} 2017, \apj, 837, 26,
  \dodoi{10.3847/1538-4357/aa5d50}

\bibitem[{{Chen} {et~al.}(2012){Chen}, {Dai}, {Kochanek}, {Chartas},
  {Blackburne}, \& {Morgan}}]{Chen2012}
{Chen}, B., {Dai}, X., {Kochanek}, C.~S., {et~al.} 2012, \apj, 755, 24,
  \dodoi{10.1088/0004-637X/755/1/24}

\bibitem[{{Cornachione} \& {Morgan}(2020)}]{Cornachione2020a}
{Cornachione}, M.~A., \& {Morgan}, C.~W. 2020, \apj, 895, 93,
  \dodoi{10.3847/1538-4357/ab8aed}

\bibitem[{{Cornachione} {et~al.}(2020){Cornachione}, {Morgan}, {Burger},
  {Shalyapin}, {Goicoechea}, {Vrba}, {Dahm}, \& {Tilleman}}]{Cornachione2020b}
{Cornachione}, M.~A., {Morgan}, C.~W., {Burger}, H.~R., {et~al.} 2020, \apj,
  905, 7, \dodoi{10.3847/1538-4357/abc25d}

\bibitem[{{Dai} {et~al.}(2003){Dai}, {Chartas}, {Agol}, {Bautz}, \&
  {Garmire}}]{Dai2003}
{Dai}, X., {Chartas}, G., {Agol}, E., {Bautz}, M.~W., \& {Garmire}, G.~P. 2003,
  \apj, 589, 100, \dodoi{10.1086/374548}

\bibitem[{{Dai} \& {Kochanek}(2009)}]{Dai2009}
{Dai}, X., \& {Kochanek}, C.~S. 2009, \apj, 692, 677,
  \dodoi{10.1088/0004-637X/692/1/677}

\bibitem[{{Dai} {et~al.}(2010){Dai}, {Kochanek}, {Chartas}, {Koz{\l}owski},
  {Morgan}, {Garmire}, \& {Agol}}]{Dai2010}
{Dai}, X., {Kochanek}, C.~S., {Chartas}, G., {et~al.} 2010, \apj, 709, 278,
  \dodoi{10.1088/0004-637X/709/1/278}

\bibitem[{{Dai} {et~al.}(2019){Dai}, {Steele}, {Guerras}, {Morgan}, \&
  {Chen}}]{Dai2019}
{Dai}, X., {Steele}, S., {Guerras}, E., {Morgan}, C.~W., \& {Chen}, B. 2019,
  \apj, 879, 35, \dodoi{10.3847/1538-4357/ab1d56}

\bibitem[{{Dogruel} {et~al.}(2020){Dogruel}, {Dai}, {Guerras}, {Cornachione},
  \& {Morgan}}]{Dogruel2020}
{Dogruel}, M.~B., {Dai}, X., {Guerras}, E., {Cornachione}, M., \& {Morgan},
  C.~W. 2020, \apj, 894, 153, \dodoi{10.3847/1538-4357/ab879b}

\bibitem[{{Fabian} {et~al.}(2015){Fabian}, {Lohfink}, {Kara}, {Parker},
  {Vasudevan}, \& {Reynolds}}]{Fabian2015}
{Fabian}, A.~C., {Lohfink}, A., {Kara}, E., {et~al.} 2015, \mnras, 451, 4375,
  \dodoi{10.1093/mnras/stv1218}

\bibitem[{{Frederick} {et~al.}(2018){Frederick}, {Kara}, {Reynolds}, {Pinto},
  \& {Fabian}}]{Frederick2018}
{Frederick}, S., {Kara}, E., {Reynolds}, C., {Pinto}, C., \& {Fabian}, A. 2018,
  \apj, 867, 67, \dodoi{10.3847/1538-4357/aae306}

\bibitem[{{Fruscione} {et~al.}(2026{\natexlab{a}}){Fruscione}, {McDowell},
  {Burke}, {Cresitello-Dittmar}, {Evans}, {Evans}, {Glotfelty}, {G{\"u}nther},
  {Huenemoerder}, {Joye}, \& et~al.}]{CIAO2026}
{Fruscione}, A., {McDowell}, J.~C., {Burke}, D.~J., {et~al.}
  2026{\natexlab{a}}, \apj, 1005, 116, \dodoi{10.3847/1538-4357/ae6db4}

\bibitem[{{Fruscione} {et~al.}(2026{\natexlab{b}}){Fruscione}, {McDowell},
  {Burke}, {Cresitello-Dittmar}, {Evans}, {Evans}, {Glotfelty}, {G{\"u}nther},
  {Huenemoerder}, {Joye}, \& et~al.}]{Fruscione2026}
---. 2026{\natexlab{b}}, \apj, 1005, 116, \dodoi{10.3847/1538-4357/ae6db4}

\bibitem[{{Garmire} {et~al.}(2003){Garmire}, {Bautz}, {Ford}, {Nousek}, \&
  {Ricker}}]{Garmire2003}
{Garmire}, G.~P., {Bautz}, M.~W., {Ford}, P.~G., {Nousek}, J.~A., \& {Ricker},
  George~R., J. 2003, in Society of Photo-Optical Instrumentation Engineers
  (SPIE) Conference Series, Vol. 4851, X-Ray and Gamma-Ray Telescopes and
  Instruments for Astronomy., ed. J.~E. {Truemper} \& H.~D. {Tananbaum},
  28--44, \dodoi{10.1117/12.461599}

\bibitem[{{Ghisellini} {et~al.}(2004){Ghisellini}, {Haardt}, \&
  {Matt}}]{Ghisellini2004}
{Ghisellini}, G., {Haardt}, F., \& {Matt}, G. 2004, \aap, 413, 535,
  \dodoi{10.1051/0004-6361:20031562}

\bibitem[{{Goicoechea} \& {Shalyapin}(2016)}]{Goicoechea2016}
{Goicoechea}, L.~J., \& {Shalyapin}, V.~N. 2016, \aap, 596, A77,
  \dodoi{10.1051/0004-6361/201628790}

\bibitem[{{Gould}(2000)}]{Gould2000}
{Gould}, A. 2000, \apj, 535, 928, \dodoi{10.1086/308865}

\bibitem[{{Haardt} \& {Maraschi}(1991)}]{Haardt1991}
{Haardt}, F., \& {Maraschi}, L. 1991, \apjl, 380, L51, \dodoi{10.1086/186171}

\bibitem[{{Haardt} \& {Maraschi}(1993)}]{Haardt1993}
---. 1993, \apj, 413, 507, \dodoi{10.1086/173020}

\bibitem[{{Hawley} \& {Balbus}(2002)}]{Hawley2002}
{Hawley}, J.~F., \& {Balbus}, S.~A. 2002, \apj, 573, 738,
  \dodoi{10.1086/340765}

\bibitem[{{HI4PI Collaboration} {et~al.}(2016){HI4PI Collaboration}, {Ben
  Bekhti}, {Fl{\"o}er}, {Keller}, {Kerp}, {Lenz}, {Winkel}, {Bailin},
  {Calabretta}, {Dedes}, {Ford}, {Gibson}, {Haud}, {Janowiecki}, {Kalberla},
  {Lockman}, {McClure-Griffiths}, {Murphy}, {Nakanishi}, {Pisano}, \&
  {Staveley-Smith}}]{HI4PI2016}
{HI4PI Collaboration}, {Ben Bekhti}, N., {Fl{\"o}er}, L., {et~al.} 2016, \aap,
  594, A116, \dodoi{10.1051/0004-6361/201629178}

\bibitem[{{Hirose} {et~al.}(2004){Hirose}, {Krolik}, {De Villiers}, \&
  {Hawley}}]{Hirose2004}
{Hirose}, S., {Krolik}, J.~H., {De Villiers}, J.-P., \& {Hawley}, J.~F. 2004,
  \apj, 606, 1083, \dodoi{10.1086/383184}

\bibitem[{Hunter(2007)}]{matplotlib}
Hunter, J.~D. 2007, Computing in Science \& Engineering, 9, 90,
  \dodoi{10.1109/MCSE.2007.55}

\bibitem[{{Inada} {et~al.}(2009){Inada}, {Oguri}, {Shin}, {Kayo}, {Strauss},
  {Morokuma}, {Schneider}, {Becker}, {Bahcall}, \& {York}}]{Inada2009}
{Inada}, N., {Oguri}, M., {Shin}, M.-S., {et~al.} 2009, \aj, 137, 4118,
  \dodoi{10.1088/0004-6256/137/5/4118}

\bibitem[{{Jim{\'e}nez-Vicente} {et~al.}(2014){Jim{\'e}nez-Vicente},
  {Mediavilla}, {Kochanek}, {Mu{\~n}oz}, {Motta}, {Falco}, \&
  {Mosquera}}]{Jimenez2014a}
{Jim{\'e}nez-Vicente}, J., {Mediavilla}, E., {Kochanek}, C.~S., {et~al.} 2014,
  \apj, 783, 47, \dodoi{10.1088/0004-637X/783/1/47}

\bibitem[{{Jovanovi{\'c}} {et~al.}(2008){Jovanovi{\'c}}, {Zakharov},
  {Popovi{\'c}}, \& {Petrovi{\'c}}}]{Jovanovic2008}
{Jovanovi{\'c}}, P., {Zakharov}, A.~F., {Popovi{\'c}}, L.~{\v{C}}., \&
  {Petrovi{\'c}}, T. 2008, \mnras, 386, 397,
  \dodoi{10.1111/j.1365-2966.2008.13036.x}

\bibitem[{{Keeton}(2001)}]{Keeton2001}
{Keeton}, C.~R. 2001, arXiv e-prints, astro.
\newblock \doarXiv{astro-ph/0102340}

\bibitem[{{Kochanek}(2004)}]{Kochanek2004}
{Kochanek}, C.~S. 2004, \apj, 605, 58, \dodoi{10.1086/382180}

\bibitem[{{MacLeod} {et~al.}(2015){MacLeod}, {Morgan}, {Mosquera}, {Kochanek},
  {Tewes}, {Courbin}, {Meylan}, {Chen}, {Dai}, \& {Chartas}}]{Macleod2015}
{MacLeod}, C.~L., {Morgan}, C.~W., {Mosquera}, A., {et~al.} 2015, \apj, 806,
  258, \dodoi{10.1088/0004-637X/806/2/258}

\bibitem[{{Maraschi} \& {Tavecchio}(2003)}]{Maraschi2003}
{Maraschi}, L., \& {Tavecchio}, F. 2003, \apj, 593, 667, \dodoi{10.1086/342118}

\bibitem[{{Martocchia} {et~al.}(2002){Martocchia}, {Matt}, {Karas}, {Belloni},
  \& {Feroci}}]{Martocchia2002}
{Martocchia}, A., {Matt}, G., {Karas}, V., {Belloni}, T., \& {Feroci}, M. 2002,
  \aap, 387, 215, \dodoi{10.1051/0004-6361:20020359}

\bibitem[{{McLeod} {et~al.}(1998){McLeod}, {Bernstein}, {Rieke}, \&
  {Weedman}}]{McLeod1998}
{McLeod}, B.~A., {Bernstein}, G.~M., {Rieke}, M.~J., \& {Weedman}, D.~W. 1998,
  \aj, 115, 1377, \dodoi{10.1086/300285}

\bibitem[{{Morgan} {et~al.}(2018){Morgan}, {Hyer}, {Bonvin}, {Mosquera},
  {Cornachione}, {Courbin}, {Kochanek}, \& {Falco}}]{Morgan2018}
{Morgan}, C.~W., {Hyer}, G.~E., {Bonvin}, V., {et~al.} 2018, \apj, 869, 106,
  \dodoi{10.3847/1538-4357/aaed3e}

\bibitem[{{Morgan} {et~al.}(2008){Morgan}, {Kochanek}, {Dai}, {Morgan}, \&
  {Falco}}]{Morgan2008}
{Morgan}, C.~W., {Kochanek}, C.~S., {Dai}, X., {Morgan}, N.~D., \& {Falco},
  E.~E. 2008, \apj, 689, 755, \dodoi{10.1086/592767}

\bibitem[{{Morgan} {et~al.}(2010){Morgan}, {Kochanek}, {Morgan}, \&
  {Falco}}]{Morgan2010}
{Morgan}, C.~W., {Kochanek}, C.~S., {Morgan}, N.~D., \& {Falco}, E.~E. 2010,
  \apj, 712, 1129, \dodoi{10.1088/0004-637X/712/2/1129}

\bibitem[{{Morgan} {et~al.}(2012){Morgan}, {Hainline}, {Chen}, {Tewes},
  {Kochanek}, {Dai}, {Kozlowski}, {Blackburne}, {Mosquera}, {Chartas},
  {Courbin}, \& {Meylan}}]{Morgan2012}
{Morgan}, C.~W., {Hainline}, L.~J., {Chen}, B., {et~al.} 2012, \apj, 756, 52,
  \dodoi{10.1088/0004-637X/756/1/52}

\bibitem[{{Mortonson} {et~al.}(2005){Mortonson}, {Schechter}, \&
  {Wambsganss}}]{Mortonson2005}
{Mortonson}, M.~J., {Schechter}, P.~L., \& {Wambsganss}, J. 2005, \apj, 628,
  594, \dodoi{10.1086/431195}

\bibitem[{{Mosquera} \& {Kochanek}(2011)}]{Mosquera2011}
{Mosquera}, A.~M., \& {Kochanek}, C.~S. 2011, \apj, 738, 96,
  \dodoi{10.1088/0004-637X/738/1/96}

\bibitem[{{Mosquera} {et~al.}(2013){Mosquera}, {Kochanek}, {Chen}, {Dai},
  {Blackburne}, \& {Chartas}}]{Mosquera2013}
{Mosquera}, A.~M., {Kochanek}, C.~S., {Chen}, B., {et~al.} 2013, \apj, 769, 53,
  \dodoi{10.1088/0004-637X/769/1/53}

\bibitem[{{Navarro} {et~al.}(1996){Navarro}, {Frenk}, \& {White}}]{Navarro1996}
{Navarro}, J.~F., {Frenk}, C.~S., \& {White}, S. D.~M. 1996, \apj, 462, 563,
  \dodoi{10.1086/177173}

\bibitem[{{Nayakshin} {et~al.}(2004){Nayakshin}, {Cuadra}, \&
  {Sunyaev}}]{Nayakshin2004}
{Nayakshin}, S., {Cuadra}, J., \& {Sunyaev}, R. 2004, \aap, 413, 173,
  \dodoi{10.1051/0004-6361:20031537}

\bibitem[{{Popovi{\'c}} {et~al.}(2006){Popovi{\'c}}, {Jovanovi{\'c}},
  {Mediavilla}, {Zakharov}, {Abajas}, {Mu{\~n}oz}, \& {Chartas}}]{Popovic2006}
{Popovi{\'c}}, L.~{\v{C}}., {Jovanovi{\'c}}, P., {Mediavilla}, E., {et~al.}
  2006, \apj, 637, 620, \dodoi{10.1086/498558}

\bibitem[{{Reis} {et~al.}(2014){Reis}, {Reynolds}, {Miller}, \&
  {Walton}}]{Reis2014}
{Reis}, R.~C., {Reynolds}, M.~T., {Miller}, J.~M., \& {Walton}, D.~J. 2014,
  \nat, 507, 207, \dodoi{10.1038/nature13031}

\bibitem[{{Reynolds} {et~al.}(2014){Reynolds}, {Walton}, {Miller}, \&
  {Reis}}]{Reynolds2014}
{Reynolds}, M.~T., {Walton}, D.~J., {Miller}, J.~M., \& {Reis}, R.~C. 2014,
  \apjl, 792, L19, \dodoi{10.1088/2041-8205/792/1/L19}

\bibitem[{{Shakura} \& {Sunyaev}(1973)}]{Shakura1973}
{Shakura}, N.~I., \& {Sunyaev}, R.~A. 1973, \aap, 500, 33

\bibitem[{{Shalyapin} \& {Goicoechea}(2014)}]{Shalyapin2014}
{Shalyapin}, V.~N., \& {Goicoechea}, L.~J. 2014, \aap, 568, A116,
  \dodoi{10.1051/0004-6361/201323360}

\bibitem[{{Shalyapin} {et~al.}(2021){Shalyapin}, {Goicoechea}, {Morgan},
  {Cornachione}, \& {Sergeyev}}]{Shalyapin2021}
{Shalyapin}, V.~N., {Goicoechea}, L.~J., {Morgan}, C.~W., {Cornachione}, M.~A.,
  \& {Sergeyev}, A.~V. 2021, \aap, 646, A165,
  \dodoi{10.1051/0004-6361/202038770}

\bibitem[{{Tinker} {et~al.}(2012){Tinker}, {Sheldon}, {Wechsler}, {Becker},
  {Rozo}, {Zu}, {Weinberg}, {Zehavi}, {Blanton}, {Busha}, \&
  {Koester}}]{Tinker2012}
{Tinker}, J.~L., {Sheldon}, E.~S., {Wechsler}, R.~H., {et~al.} 2012, \apj, 745,
  16, \dodoi{10.1088/0004-637X/745/1/16}

\bibitem[{{Vernardos} \& {Tsagkatakis}(2019)}]{Vernardos2019}
{Vernardos}, G., \& {Tsagkatakis}, G. 2019, \mnras, 486, 1944,
  \dodoi{10.1093/mnras/stz868}

\bibitem[{{Wambsganss} {et~al.}(1992){Wambsganss}, {Witt}, \&
  {Schneider}}]{Wambsganss1992}
{Wambsganss}, J., {Witt}, H.~J., \& {Schneider}, P. 1992, \aap, 258, 591

\end{thebibliography}

\end{document}